\documentclass[journal, twoside, final]{IEEEtran}
\usepackage{amssymb, amsmath, amsthm, amsfonts}
\usepackage{bm, color}
\usepackage{booktabs}
\usepackage{algorithm}
\usepackage{algorithmic}
\usepackage{url}
\usepackage{graphicx}
\usepackage{longtable}
\usepackage{makecell}
\usepackage{cite}
\usepackage{color}
\usepackage{dsfont}
\usepackage{datetime}
\usepackage{float}
\usepackage{subfigure} 

\newtheorem{remark}{Remark}

\newtheorem{prop}{Proposition}
\allowdisplaybreaks[4]

\usepackage{threeparttable}

\begin{document}
	\title{Cooperative Beamforming for Wireless Fronthaul and Access Links in Ultra-Dense C-RANs with SWIPT: A First-Order Approach}
	\author{Fangqing Tan, Peiran Wu, \IEEEmembership{Member,~IEEE}, Yik-Chung Wu, \IEEEmembership{Senior Member,~IEEE}, \\ and Minghua Xia, \IEEEmembership{Senior Member,~IEEE}
		
\thanks{Manuscript received December 30, 2020; revised May 12, 2021; accepted May 25, 2021. This work was supported in part by the Key-Area Research and Development Program of Guangdong Province under Grant 2018B010114001, in part by the National Natural Science Foundation of China under Grants 62001521, 61801526, and U2001213, in part by the Guangxi Natural Science Foundation under Grants 2018GXNSFBA138034 and AD18281052, in part by the Fund of Key Laboratory of Cognitive Radio and Information Processing, Ministry of Education, China, under Grant CRKL180103, in part by the Fundamental Research Funds for the Central Universities under Grant 191gjc04, and in part by the China Postdoctoral Science Foundation under Grant 2019M653177. (\textit{Corresponding author: Minghua Xia}.)
	
	Fangqing Tan, Peiran Wu, and Minghua Xia are with the School of Electronics and Information Technology, Sun Yat-sen University, Guangzhou 510006, China. Fangqing Tan is also with the Key Laboratory of Cognitive Radio and Information Processing, Guilin University of Electronic Technology, Guilin 541004, China. Peiran Wu and Minghua Xia are also with the Southern Marine Science and Engineering Guangdong Laboratory, Zhuhai 519082, China (e-mail: \{tanfq, wupr3, xiamingh\}@mail.sysu.edu.cn).}

\thanks{Yik-Chung Wu is with the Department of Electrical and Electronic Engineering, The University of Hong Kong, Hong Kong (e-mail: ycwu@eee.hku.hk).}

\thanks{Color versions of one or more of the figures in this article are available online at https://ieeexplore.ieee.org.
	
	Digital Object Identifier XXX}
}

\markboth{IEEE Journal of Selected Topics in Signal Processing} {Tan \MakeLowercase{\textit{et al.}}: Cooperative Beamforming for Wireless Fronthaul and Access Links in Ultra-Dense C-RANs with SWIPT}

\maketitle

\IEEEpubid{\begin{minipage}{\textwidth} \ \\[35pt] \centering 0733-8716 \copyright\ 2021 IEEE. Personal use is permitted, but republication/redistribution requires IEEE permission. \\
	See \url{https://www.ieee.org/publications/rights/index.html} for more information.\end{minipage}}
	
	\begin{abstract}
		\noindent This work studies multigroup multicasting transmission in cloud radio access networks (C-RANs) with simultaneous wireless information and power transfer, where densely packed remote radio heads (RRHs) cooperatively provide information and energy services for information users (IUs) and energy users (EUs), respectively. To maximize the weighted sum rate (WSR) of information services while satisfying the energy harvesting levels at EUs, an optimization of joint beamforming design for the fronthaul and access links is formulated, which is however neither smooth nor convex and is indeed NP-hard. To tackle this difficulty, the smooth and successive convex approximations are used to transform the original problem into a sequence of convex problems, and two first-order algorithms are developed to find the initial feasible point and the nearly optimal solution, respectively. Moreover, an accelerated algorithm is designed to improve the convergence speed by exploiting both Nesterov and heavy-ball momentums. Numerical results demonstrate that the proposed first-order algorithms achieve almost the same WSR as that of traditional second-order approaches yet with much lower computational complexity, and the proposed scheme outperforms state-of-the-art competing schemes in terms of WSR.
	\end{abstract}
	
	
	\vspace{-10pt}
	\section{Introduction}
	With the explosive development of the Internet-of-Things (IoTs), billions of low-power consumption devices (e.g., smart terminals, sensors and wearables) are deployed in various smart applications. Such massive networks, while providing ubiquitous communication connectivity, require perpetual energy supply \cite{9170653}. To this end, simultaneous wireless information and power transfer (SWIPT), where the received (Rx) signals are exploited for information decoding and energy harvesting, has been widely accepted as a promising technology for power-limited IoT networks. In practice, however, a typical energy user (EU) such as humidity sensor requires much higher energy for its operation than for information users (IUs), due to the different sensitivities between energy harvesting circuitry and information decoding circuitry. In addition, severe channel attenuation leads to low power transfer efficiency and this constitutes a major bottleneck that hinders deploying massive energy-harvesting IoTs \cite{7070727}. Therefore, further improving the energy transfer efficiency is paramount in beyond fifth generation (5G) wireless networks \cite{8476597}.
	
	On the other hand, cloud radio access networks (C-RANs), where low-power remote radio heads (RRHs) are densely deployed and connected to a pool of baseband processing units (BBUs) at the computation center via fronthaul links, has been deemed as a prospective network architecture, due to its potential for achieving substantial spectral efficiency and energy efficiency (EE) \cite{7018201}. Due to the dense arrangement, it is more likely that users are close to one or a few RRHs, which will render higher degrees of macro-diversity and lower path loss and is beneficial for energy harvesting. Consequently, the integration of SWIPT and C-RANs is of great practical significance for enhancing wireless power transfer and realizing the envisioned battery-free IoT networks in the future \cite{7462481}. 
	
	In C-RANs, the capacity of fronthaul connecting the computation center to RRHs determines the data rate and coverage area of the network. In general, wired fronthaul links are desired because of its large capacity. However, albeit large, the fixed capacity makes wired fronthaul links inflexible to cope with dynamic traffic and/or unplanned traffic in the future.  Also, if wired fronthaul links are overdesigned with high bandwidth (e.g., cables or optical fibers), it may be impractical for ultra-dense C-RANs due to its extremely high deployment cost, especially in urban areas \cite{7128726}. Therefore, as an alternative, {\it wireless fronthaul} becomes a viable  option in C-RANs, thanks to its flexibility, scalability, and low deployment cost \cite{6588655}. 
	
	Some innovative efforts have been devoted to the fronthaul and access links design for C-RANs \cite{7924402,7573000,7809091,8283646,8830409,9001081,8976409}. Specifically, the authors of \cite{7924402} proposed an effective beamforming design for C-RANs by accounting for EE and wired/wireless fronthaul cost. The work \cite{7573000} iteratively optimized the fronthaul compression and hybrid beamforming for maximizing the weighted sum rate (WSR) and EE in C-RANs. Later, the work \cite{7809091} developed a joint resource allocation in ultra-dense C-RANs, by considering joint mmWave fronthaul and wireless access transmission optimization. Two algorithms were developed to design wireless fronthaul and access links for C-RANs in \cite{8283646}, by using the difference of convex programming and successive convex approximation (SCA) schemes. In addition, the work \cite{8830409} proposed to employ hybrid RF/FSO systems for wireless fronthauling of C-RANs, and the RF transmission time allocated to the multiple-access
	and fronthaul links was adaptively optimized. More recently, a secure beamforming was designed in \cite{9001081} for the base station (BS)-cooperation-aided mmWave C-RANs with a microwave multicast fronthaul. The work \cite{8976409} made a joint cache allocation and beamforming design for maximizing the content downloading sum-rate in C-RANs with multi-cluster multicast wireless backhaul. To make a balance between the system performance and the fronthaul overhead, user-centric coordinated transmission schemes are proposed, in which only a part of RRHs form a coordination cluster and perform joint beamforming, and only the coordinated RRHs in the same cluster need to share the users’ data information. The up-to-date survey \cite{ammar2021usercentric} made a comprehensive review of the theories and techniques devoted to user-centric cell-free networks. A user-centric BS clustering and beamformer design problem was jointly studied \cite{6415394}, and an efficient algorithm was proposed by iteratively solving a sequence of group LASSO problem. On the other hand, to reduce the computational complexity and channel estimation overhead for coordination in C-RANs, the work \cite{7314952} developed a unified theoretical framework for dynamic clustering by exploiting the near-sparsity of large channel matrices. In addition, the content-centric BS clustering was developed to reduce the traffic load over the fronthaul links, by using the content diversity \cite{7488289}. 
	
	While the above works focus on exploiting cooperative beamforming for enhancing the information transmission performance in C-RANs, the high beamforming gain achieved by cooperative beamforming is also appealing for wireless power transfer \cite{7106496,8540005,8948235,7942080,7565627}. By leveraging cooperative beamforming, C-RANs can help
	compensate the high RF signal attenuation over long distance and thereby achieve a higher energy transfer efficiency. To reap this benefit, the work \cite{7106496} studied resource allocation algorithm for secure information and renewable green energy transmission in distributed antenna systems. Later, \cite{8540005} studied signal processing strategies for downlink and uplink of C-RANs with SWIPT. Furthermore, the authors of \cite{8948235} jointly designed beamforming and power splitting in C-RANs with multicast fronthaul. For full-duplex C-RANs, a joint transceiver design was developed in \cite{7942080} to minimize the total power consumption. Recently, to strike an optimum balance among the total power consumption in fronthaul links, the authors of \cite{7565627} proposed two joint real-time resource allocation and energy trading strategies. However, a common feature of all the aforementioned works is the use of a second-order interior-point method whose complexity order is ${\cal O}(N^{3})$ with $N$ being the problem size, limiting previous studies to small/medium-scale C-RANs with no more than dozens of users and RRHs, which is not suitable for ultra-dense C-RANs.
	
	Motivated by the above observations, this paper develops a low-complexity cooperative beamforming for ultra-dense C-RANs with SWIPT, where a large number of RRHs jointly provide information and energy services for massive IUs and EUs, respectively. For efficient data transmission, the network provides not only multicast services but also broadcast services. To sufficiently leverage  wireless fronthaul and SWIPT, a cooperative beamforming for fronthaul and access links is designed via a first-order algorithm.  The network under study is of practical importance for massive access applications in, e.g.,  smart cities and intelligent industry, where high-density machine-type communications are indispensable. In particular, the main contributions of this paper are summarized as follows. 
\begin{itemize}
	\item A cooperative beamforming for fronthaul and access links is designed to maximize the WSR of information services while satisfying the energy harvesting levels at EUs. Specifically, the original nonsmooth and nonconvex optimization problem is firstly transformed into a sequence of convex problems by using the smooth approximation and SCA techniques. Instead of directly solving each convexified problem with the interior-point method, a strong convex upper bound of the objective function is further constructed and, then, a first-order algorithm is designed to solve each SCA subproblem in the dual domain. The designed first-order algorithm obtains almost the same WSR as traditional second-order algorithms yet with much lower computational complexity.
	\item To improve the convergence speed of the designed first-order algorithm, an accelerated algorithm is further developed by jointly exploiting Nesterov and heavy-ball momentums. As each iterative update in the accelerated algorithm is more aggressive than the conventional gradient step, it converges more than twice as fast as the first-order algorithm, without loss of WSR performance.
	\item To identify a feasible initial point, the original feasibility problem is equivalently transformed into a nonconvex optimization problem with two simple and independent sets of constraints. Then, another first-order algorithm is developed to efficiently solve the transformed optimization by using the stochastic subgradient descent method. The convergence of the proposed algorithms are guaranteed and extensive Monte-Carlo simulation results demonstrate their effectiveness.
\end{itemize}
	
	To detail the aforementioned contributions, the rest of this paper is organized as follows.  Section~\ref{Sec_sys_m} describes the system model. Section~\ref{Sec_EE_RA} formulates the WSR maximization problem and designs a first-order algorithm and its accelerated version. Section~\ref{Sec_ffp} develops another first-order algorithm to find a feasible initial point. Simulation results are discussed in Section~\ref{Sec_sim} and, finally, Section~\ref{Sec_con} concludes the paper.
	
	{\it Notation:} Vectors and matrices are denoted by lower- and upper-case letters in boldface, respectively. The operators $|\cdot|$, $\|\cdot\|_{F}$, $(\cdot)^{\rm T}$, $(\cdot)^{\rm H}$, ${\rm Tr}(\cdot)$ and $(\cdot)^{-1}$ indicate the determinant, Frobenius norm, transpose, conjugate transpose, trace and inverse of a matrix, respectively. The operator $\Re\{\cdot\}$ takes the real part of a complex number and ${\mathbb E}[\cdot]$ gives the statistical expectation of a random variable. The symbol ${\rm diag}({\bm x})$ constructs a diagonal matrix with entries specified by ${\bm x}$, and $\nabla f({\bm x})$ stands for the gradient of $f({\bm x})$. Finally, $[x]^{+} \triangleq \max\{0, x\}$, and ${\bm 0}_{M}$ and ${\bm 1}_{M}$ refer to the all-zero and all-one vectors of length $M$, respectively.
	
	\section{System Model}\label{Sec_sys_m}

\begin{figure}[!t]
		\includegraphics[width=3.0in]{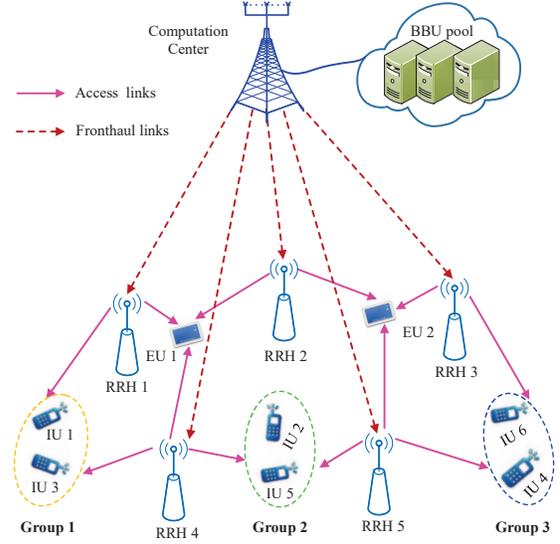}
		\centering{}
		\vspace{-5pt}	
		\caption{The system model of multigroup multicasting C-RANs with SWIPT.}
		\label{Fig_1}
	\end{figure}

	As illustrated in Fig.~\ref{Fig_1}, we consider the downlink of a C-RAN system with SWIPT, consisting of a computation center, a BBU pool, $N$ RRHs, $K$ IUs, and $Q$ EUs, where the sets of RRHs, IUs and EUs are denoted by ${\cal N} \triangleq \{1, \cdots, N\}$, ${\cal K} \triangleq \{1, \cdots, K\}$ and ${\cal Q} \triangleq \{1, \cdots, Q\}$, respectively. Let $L$, $M$, ${T}_{\rm I}$ and ${T}_{\rm E}$ be the numbers of antennas at the computation center, each RRH, IU and EU, respectively. For practical purposes and without loss of generality, it is assumed $L\gg M \geq T_{\rm I}, T_{\rm E} \geq 1$. The fronthaul links (i.e., the links from the computation center to RRHs) and access links (i.e., the links from RRHs to IUs/EUs) are supposed to work in exclusive frequency or time domain, yielding no co-channel interference between them. 
	
	\subsection{Access Links}
	According to their different multicast service subscriptions, all IUs are divided into $G$ multicast groups ${\cal K}_{1}, \cdots, {\cal K}_{G}$, where ${\cal K}_{g}$ is the set of IUs in group $g$, $\forall g\in{\cal G}\triangleq\{1, \cdots, G\}$ with $1 \leq G \leq K$.  It is assumed that each IU belongs to one and only one group, that is, $\mathop{\cup}_{g \in {\cal G}}{\cal K}_{g} = {\cal K}$ and ${\cal K}_{i}\cap{\cal K}_{g} = \emptyset, \forall i \neq g \in {\cal G}$. In addition to the multicast services, all IUs receive a common broadcast service as well.
	
	Let ${\bm s}_{0}\in{\mathbb C}^{d_{\rm a}\times 1}$ be the broadcast message for all IUs and ${\bm s}_{g} \in {\mathbb C}^{d_{a} \times 1}, \forall g \in {\cal G}$ be the multicast message for the IUs in group $g$, where $d_{\rm a}$ is the number of data streams destined for each IU, with $1 \leq d_{\rm a}\leq T_{\rm I}$ and ${\mathbb E}[\|{\bm s}_{0}\|_{2}^{2}] = {\mathbb E}[\|{\bm s}_{g}\|_{2}^{2}] = 1$. Then, the transmit (Tx) signal from RRH $n$ is given by
\begin{equation}\label{TX}
	\bm{x}_{n} =\sum_{j\in0\cup{\cal G}}\bm{V}_{n,j}{\bm s}_{j},\ \forall n\in{\cal N},
\end{equation}
where $\bm{V}_{n,j}\in\mathbb{C}^{M\times d_{\rm a}}$ is the beamforming matrix at RRH $n$ pertaining to message ${\bm s}_{j}$. Accordingly, the Rx signal at IU $k$ can be expressed as
	\begin{equation}\label{RX_k}
		{\bm y}_{k}^{\rm I} =\bm{H}_{k}\bm{V}_{0}{\bm s}_{0}+\bm{H}_{k}\bm{V}_{g_{k}}{\bm s}_{g_{k}}
		+\bm{H}_{k}\sum_{g \in {\cal G}\setminus g_{k}}\bm{V}_{g}{\bm s}_{g}+{\bm n}_{k},
	\end{equation}
	where $g_{k}\in{\cal G}$ is the group index to which IU $k$ belongs; $\bm{H}_{k}\in\mathbb{C}^{T_{\rm I}\times MN}$ is the network-wide channel matrix between all RRHs and IU $k$; $\bm{V}_{j}=\left[\bm{V}_{1,j}^{\rm T},\cdots,\bm{V}_{N,j}^{\rm T}\right]^{\rm T}\in\mathbb{C}^{MN\times d_{\rm a}}$ stands for the network-wide beamforming matrix corresponding to ${\bm s}_{j}$; and ${\bm n}_{k}\sim{\cal CN}\left(0,\delta_{\rm a}^{2}{\bm I}_{T_{\rm I}}\right)$ means an additive white Gaussian noise (AWGN) at IU $k$. Next, we compute the achievable data rate of IUs and the harvested energy of EUs.

	\subsubsection{Information Service}
	Regarding the information decoding procedure, IU $k$ decodes both ${\bm s}_{0}$ and ${\bm s}_{g_{k}}$  using successive interference cancellation strategy. In particular, as the 
	broadcast message has a higher priority and/or higher Tx power than multicast messages due to its wider coverage, the broadcast message ${\bm s}_{0}$ is first decoded by treating all multicast signals as Gaussian noise and then, after subtracting the broadcast message from the Rx signal, each IU decodes its own multicast message. In view of \eqref{RX_k}, the achievable data rate for the broadcast and multicast services at IU $k$ can be readily computed as 
	\begin{align}
		R_{{\rm B},k}&= \log_{2}\left|\bm{I}_{d_{\rm a}}+\bm{V}_{0}^{\rm H}\bm{H}_{k}^{\rm H}{\bm J}_{{\rm b},k}^{-1}\bm{H}_{k}\bm{V}_{0}\right|,\ \forall k\in{\cal K},\label{Rc_K} \\
		R_{{\rm M},k}&= \log_{2}\left|\bm{I}_{d_{\rm a}}+\bm{V}_{g_{k}}^{\rm H}\bm{H}_{k}^{\rm H}{\bm J}_{{\rm m},k}^{-1}\bm{H}_{k}\bm{V}_{g_{k}}\right|,\ \forall k\in{\mathcal K},\label{Rp_K}
	\end{align}
	respectively, where  ${\bm J}_{{\rm b},k} \triangleq \sum_{g\in{\cal G}}\bm{H}_{k}\bm{V}_{g}\bm{V}_{g}^{\rm H}\bm{H}_{k}^{\rm H}+\delta_{a}^{2}\bm{I}_{T_{\rm I}}$ and ${\bm J}_{{\rm m},k} \triangleq \sum_{g\in{\cal G}\setminus g_{k}}\bm{H}_{k}\bm{V}_{g}\bm{V}_{g}^{\rm H}\bm{H}_{k}^{\rm H}+\delta_{a}^{2}\bm{I}_{T_{\rm I}}$ indicate the interference-plus-noise covariance matrixes for the broadcast and multicast services at IU $k$, respectively. Finally, with \eqref{Rc_K}-\eqref{Rp_K}, the overall achievable data rate of the network for the broadcast service, denoted $R_{0}$, is determined by the minimum of all $R_{{\rm B}, k}$, $k \in {\mathcal K}$, while the achievable data rate of the network for multicast service for group $g$, denoted $R_{g}, g \in {\cal G}$, is determined by the minimum of all $R_{{\rm M},k}$, $k \in {\cal K}_{g}$ \cite{8976409}, i.e.,
	\begin{equation}
		R_{0} = \min_{k\in{\cal K}}\left\{R_{{\rm B},k}\right\}, \
		R_{g} = \min_{k\in{\cal K}_{g}}\left\{R_{{\rm M},k}\right\}, \forall g\in{\cal G}. \label{cpp_1} 
	\end{equation}
	
	\subsubsection{Energy Service}
	As for wireless power transfer, the Rx signal at EU $q$ is given by
	\begin{equation} \label{yeq}
		{\bm y}_{q}^{\rm E}=\bm{F}_{q}\sum_{j\in0\cup{\cal G}}\bm{V}_{j}{\bm s}_{j}+{\bm n}_{q},\ \forall q\in{\cal Q},
	\end{equation}
	where $\bm{F}_{q} \in \mathbb{C}^{T_{\rm E}\times MN}$ denotes the network-wide channel matrix between all RRHs and EU $q$, and ${\bm n}_{q}\sim{\cal CN}\left(0, \delta_{\rm e}^{2}{\bm I}_{T_{\rm E}}\right)$ is an AWGN at EU $q$. Due to the broadcast property of wireless channels, the energy carried by information bits can be harvested at each EU. By \eqref{yeq} and ignoring the noise power at EUs, the Rx RF power of EU $q$ can be computed as
	\begin{equation}\label{RF_i}
		P_{{\rm R}, q} = \sum_{j \in 0\cup{\cal G}}{\rm Tr}\left(\bm{V}_{j}^{\rm H}\bm{F}_{q}^{\rm H}\bm{F}_{q}\bm{V}_{j}\right), \forall q \in {\cal Q}.
	\end{equation}
	Then, the harvested energy at EU $q$ is counted as $E_{{\rm R},q}={\cal F}\left(P_{{\rm R},q}\right)$, where the function ${\cal F}(\cdot)$ reflects the energy conversion process. In this paper, the practical non-linear EH model developed in \cite{8322450} is adopted:
	\begin{equation} \label{EH_q}
		{\cal F}(x) = \left[\frac{P_{\max}}{e^{-\iota_{1}P_{0}+\iota_{2}}}\left(\frac{1+e^{-\iota_{1}P_{0}+\iota_{2}}}{1+e^{-\iota_{1}x+\iota_{2}}}-1\right)\right]^{+}, 
	\end{equation}
	where the parameter $P_{0}$ denotes the harvester's sensitivity threshold and $P_{\max}$ refers to the maximum harvested power when the EH circuit is saturated, and the parameters $\iota_{1}$ and $\iota_{2}$ are used to capture the nonlinear dynamics of EH circuits. 
	
	\vspace{-7pt}\subsection{Fronthaul Links}
	After the computation center acquires data of all IUs from the BBUs pool, the requested data of associated users for each RRH are multiplexed and delivered through wireless fronthauling. Let ${\bm G}_{n}\in{\mathbb C}^{M\times L}$ denote the fronthaul channel between the computation center and RRH $n$, then the Rx signal at RRH $n$ can be expressed as 
	\begin{equation}\label{Eq-1}
		{\bm y}_{n}^{\rm F}={\bm G}_{n}{\bm U}_{n}{\bm m}_{n}+{\bm G}_{n}\sum_{\ell \in{\cal N}\setminus n}{\bm U}_{\ell}{\bm m}_{\ell}+{\bm z}_{n},\ \forall n\in{\cal N},
	\vspace{-2pt}
	\end{equation}
	where ${\bm m}_{\ell}\in{\mathbb C}^{d_{\rm f}\times 1}$ and ${\bm U}_{\ell}\in{\mathbb C}^{L\times d_{\rm f}}$ denote the normalized Tx signal (i.e., ${\mathbb E}[\|{\bm m}_{\ell}\|_{2}^{2}]=1$) and corresponding beamforming matrix at the computation center for RRH $\ell$, respectively, with $1 \leq d_{\rm f} \leq M$ being the number of independent data streams; and ${\bm z}_{n} \sim {\cal CN}({\bm 0},\sigma_{\rm f}^{2}{\bm I}_{M})$ denotes an AWGN. In light of \eqref{Eq-1}, the achievable fronthaul data rate for RRH $n$ can be computed as
	\begin{equation}
		R_{{\rm F},n}=\log_{2}\left|\bm{I}_{d_{\rm f}}+\bm{U}_{n}^{\rm H}\bm{G}_{n}^{\rm H}{\bm J}_{{\rm f},n}^{-1}\bm{G}_{n}\bm{U}_{n}\right|,\ \forall n\in{\cal N},\label{Rf_n}
	\vspace{-2pt}
	\end{equation}
	where ${\bm J}_{{\rm f},n} \triangleq \sum_{\ell\in{\cal N}\setminus n}\bm{G}_{n}\bm{U}_{\ell}\bm{U}_{\ell}^{\rm H}\bm{G}_{n}^{\rm H}+\sigma_{\rm f}^{2}\bm{I}_{M}$.
	
	Due to the limited fronthaul capacity, each RRH cannot serve all but only a part of users at a moment. In this regard, the indicator function $\mathds{1}\left\{\left\|\bm{V}_{n, j}\right\|_{F}^{2}\right\}$ is introduced to characterize the relationship between service association and the beamformer for broadcast/multicast message ${\bm s}_{j}$ at RRH $n$, that is, 
	\begin{equation}\label{PC_APP}
		\mathds{1}\left\{\left\|\bm{V}_{n, j}\right\|_{F}^{2}\right\} = 
		\left\{
		\begin{array}{rl}
			1, & \text{if } \left\|\bm{V}_{n, j} \right\|_{F}^{2} \neq 0; \\
			0, & \text{otherwise},
		\end{array}
		\right.
	\end{equation}
	where $\mathds{1}\{\left\|\bm{V}_{n, j}\right\|_{F}^{2}\} = 1$ implies that group service $j$ is associated with RRH $n$, and $\mathds{1}\{\left\|\bm{V}_{n,j}\right\|_{F}^{2}\} = 0$ otherwise. To simplify the subscript, $\left\|\bm{V}_{n,j}\right\|_{F}^{2}$ can be rewtitten as $\left\|\bm{A}_{n}\bm{V}_{j}\right\|_{F}^{2}$, where $\bm{A}_{n}\triangleq\text{diag}\left\{[{\bm 0}_{(n-1)M},{\bm 1}_{M},{\bm 0}_{(N-n)M}]\right\}$. As a result, the aggregate data rate transmitted over the fronthaul link from the computation center to RRH $n$ is bound by its fronthaul link rate $R_{{\rm F},n}$, expressed as
	\begin{equation}\label{C_fh}
		R_{{\rm A},n}=\sum_{j\in0\cup{\mathcal G}}\mathds{1}\left\{\left\|\bm{A}_{n}\bm{V}_{j}\right\|_{F}^{2}\right\}R_{j}\leq R_{{\rm F},n}\ \forall n\in{\cal N}.
	\end{equation}
	
\begin{remark}[On channel acquisition and time synchronization]
	For ease of tractability, we assume that perfect channel state information (CSI) is available at the computation center and all RRHs can precisely synchronize with one another, like \cite{NOUM_1}. In practice, the CSI can be acquired as follows. At first, all users send uplink pilots to RRHs in a time-division duplexing fashion.  After estimating its own CSI, each RRH sends it to the computation center via a  fronthaul link \cite{8247283}. As for the time synchronization, integrating global positioning system into network synchronization protocol is efficient to synchronize distant RRHs, see e.g., \cite{9183752}.
\end{remark}
	
	\section{First-Order Beamforming Design}
	\label{Sec_EE_RA}
	Now, we start with maximizing the WSR of all information services by jointly designing beamformers $\{{\bm V}_{j}\}_{j\in0\cup{\cal G}}$ and $\{{\bm U}_{n}\}_{n\in{\cal N}}$. In light of \eqref{Rc_K}-\eqref{C_fh}, the problem can be formulated as
	\begin{subequations} \label{P0}
		\begin{align}
			\max_{\mathcal V} \ & \sum_{j\in0\cup{\cal G}}\alpha_{j}R_{j}({\cal V})\label{Obj} \\
			{\rm s.t.} \  
			& \ R_{{\rm A},n}({\cal V})\leq R_{{\rm F},n}({\cal V}), \ \forall n \in \mathcal{N}, \label{C1} \\
			& \ E_{{\rm R},q}({\cal V})\geq e_{q},\ \forall q\in{\cal Q}, \label{C2} \\
			& \ P_{{\rm T},n}({\cal V})\leq p_{n}, \ \forall n \in \mathcal{N}, \label{C3}\\
			& \ \sum_{\ell\in {\cal N}}\|\bm{U}_{\ell}\|_{F}^{2}\leq p_{c}, \label{C4}
		\end{align}
	\end{subequations}
	where ${\mathcal V}\triangleq\{\{{\bm V}_{j}\}_{j\in0\cup{\cal G}},\{{\bm U}_{n}\}_{n\in{\cal N}}\}$ denotes the set of all Tx beamforming matrixes pertaining to the fronthaul and access links; $\alpha_{j}$ in \eqref{Obj} represents the priority of different group services;  \eqref{C1} is the fronthaul links capacity constraint; \eqref{C2} specifies energy constraint for EU $q$, with $e_{q}$ being the minimal energy harvested requirement for EU $q$; \eqref{C3} and \eqref{C4} impose Tx power constraints for RRH $n$ and computation center, with $p_{n}$ and $p_{c}$ being the allowable maximal Tx power at RRH $n$ and the computation center, respectively, as well as $P_{{\rm T},n}({\cal V})\triangleq\sum_{j\in 0\cup{\cal G}}\|\bm{A}_{n}\bm{V}_{j}\|_{F}^{2}$.
	
	It is not hard to identify that problem \eqref{P0} is neither smooth nor convex, and even finding a feasible point satisfying the nonconvex constraints \eqref{C1}-\eqref{C2} is NP-hard. A prevailing technique to solve this problem is various second-order interior-point methods, through off-the-shelf convex solvers. However, it is well-known that interior-point methods require computing Hessian matrices, and their extremely high computational complexity and memory requirement makes them unsuitable for ultra-dense C-RANs, where the amount of RRHs and users is very large. To tackle these challenges, in the following a low-complexity first-order algorithm and its accelerated algorithm are developed, provided that a feasible initial point is available. Afterwards, another first-order algorithm is designed to find a feasible initial point.

	\subsection{Tackling Nonsmoothness and Nonconvexity of Problem \eqref{P0}}
	We begin with settling the nonsmoothness of $R_{0}(\mathcal{V})$ and $R_{g}(\mathcal{V})$ involved in \eqref{Obj}. By introducing a set of auxiliary variables $\mathcal{R}\triangleq \left\{\mathcal{R}_{0},\{\mathcal{R}_{g}\}_{g \in {\cal G}}\right\}$, such that
	\begin{align}
		{\cal R}_{0} & \leq R_{{\rm B},k}(\mathcal{V}),\ \forall k\in{\cal K}, \label{R_ck}\\
		{\cal R}_{g}& \leq R_{{\rm M},k}(\mathcal{V}),\ \forall k \in \mathcal{K}_{g}, \label{R_gk}
	\end{align}
	the nonsmooth broadcast data-rate function $R_{0}(\mathcal{V})$ and multicast data-rate function $R_{g}(\mathcal{V})$ can be replaced by $\mathcal{R}_{0}$ and $\mathcal{R}_{g}$, respectively. However, since $R_{{\rm B}, k}(\mathcal{V})$ and $R_{{\rm M},k}(\mathcal{V})$ are nonconcave, constraints \eqref{R_ck} and \eqref{R_gk} are nonconvex.
	
	To deal with the nonconvexity of \eqref{R_ck} and \eqref{R_gk}, we construct a sequence of convex constraints to approximate \eqref{R_ck} and \eqref{R_gk} by quadratically convexifying  $R_{{\rm B},k}(\mathcal{V})$ and $R_{{\rm M},k}(\mathcal{V})$, in the following proposition.
	
	\begin{prop} \label{Proposition-1}
		Given a fixed point $\mathcal{V}^{(t)}$, the lower bounding concave quadrative functions for  $R_{{\rm B},k}(\mathcal{V})$ and $R_{{\rm M},k}(\mathcal{V})$ are defined by
		\begin{equation} \label{rate_mm}
			\bar{R}_{{\rm B},k}^{(t)}({\cal V})\triangleq \vartheta_{{\rm b},k}^{(t)}-\sum_{j\in0\cup{\cal G}}{\rm Tr}({\bm V}_{j}^{\rm H}{\bm \varXi}_{{\rm b},k}^{(t)}{\bm V}_{j})-{\Re}\{{\rm Tr}({\bm \varUpsilon}_{{\rm b},k}^{(t)}{\bm V}_{0})\},
		\end{equation}
		with
		\begin{subequations}
			\begin{align}
				\vartheta_{{\rm b},k}^{(t)} &= -{\rm Tr}\left(({\bm I}_{d_{\rm a}}-\bm{\varTheta}_{{\rm b},k}^{(t){\rm H}}{\bm H}_{k}\bm{V}_{0}^{(t)})^{-1}({\bm I}_{d_{\rm a}}+\delta_{\rm a}^{2}\bm{\varTheta}_{{\rm b},k}^{(t){\rm H}}\bm{\varTheta}_{{\rm b},k}^{(t)})\right) \nonumber \\
				&\quad {}-\log_2\left|{\bm I}_{d_{\rm a}}-\bm{\varTheta}_{{\rm b},k}^{(t){\rm H}}\bm{H}_{k}\bm{V}_{0}^{(t)}\right|+d_{\rm a}, \label{rate_f_a}\\
				\bm{\varXi}_{{\rm b},k}^{(t)} &= \bm{H}_{k}^{\rm H}\bm{\varTheta}_{{\rm b},k}^{(t)}({\bm I}_{d_{\rm a}}-\bm{\varTheta}_{{\rm b},k}^{(t){\rm H}}
				\bm{H}_{k}\bm{V}_{0}^{(t)})^{-1}\bm{\varTheta}_{{\rm b},k}^{(t){\rm H}}\bm{H}_{k}, \label{rate_f_D}\\
				\bm{\varUpsilon}_{{\rm b},k}^{(t)} &= -2({\bm I}_{d_{\rm a}}-\bm{\varTheta}_{{\rm b},k}^{(t){\rm H}}\bm{H}_{k}\bm{V}_{0}^{(t)})^{-1}
				{\bm \varTheta}_{{\rm b},k}^{(t){\rm H}}\bm{H}_{k}, \label{rate_f_B}\\
				\bm{\varTheta}_{{\rm b},k}^{(t)} &= \Big(\sum_{j\in0\cup{\cal G}}\bm{H}_{k}\bm{V}_{0}^{(t)}\bm{V}_{0}^{(t)\rm H}\bm{H}_{k}^{\rm H}
				+\delta_{\rm a}^{2}\bm{I}_{T_{\rm I}}\Big)^{-1}\bm{H}_{k}\bm{V}_{0}^{(t)},  \label{rate_f_e}
			\end{align}
		\end{subequations}
		and	
		\begin{equation} \label{rate_uu}
			\bar{R}_{{\rm M},k}^{(t)}({\cal V})\triangleq \vartheta_{{\rm m},k}^{(t)}-\sum_{g\in{\cal G}}{\rm Tr}({\bm V}_{g}^{\rm H}
			{\bm \varXi}_{{\rm m},k}^{(t)}{\bm V}_{g})-{\Re}\{{\rm Tr}({\bm \varUpsilon}_{{\rm m},k}^{(t)}{\bm V}_{g_{k}})\},
		\end{equation}
		with
		${\bm{\varXi}_{{\rm m},k}^{(t)}=\bm{H}_{k}^{\rm H}\bm{\varTheta}_{{\rm m},k}^{(t)}({\bm I}_{d_{\rm a}}-\bm{\varTheta}_{{\rm m},k}^{(t){\rm H}}\bm{H}_{k}\bm{V}_{g_{k}}^{(t)})^{-1}\bm{\varTheta}_{{\rm m},k}^{(t){\rm H}}\bm{H}_{k}}$, $\bm{\varUpsilon}_{{\rm m},k}^{(t)}=-2({\bm I}_{d_{\rm a}}-\bm{\varTheta}_{{\rm m},k}^{(t){\rm H}}\bm{H}_{k}\bm{V}_{g_k}^{(t)})^{-1}{\bm \varTheta}_{{\rm m},k}^{(t){\rm H}}\bm{H}_{k}$, 
		$\!\bm{\varTheta}_{{\rm m},k}^{(t)}=\big(\sum_{g\in{\cal G}}\bm{H}_{k}\bm{V}_{g}^{(t)}\bm{V}_{g}^{(t)\rm H}\bm{H}_{k}^{\rm H}+\delta_{\rm a}^{2}\bm{I}_{T_{\rm I}}\big)^{-1}\bm{H}_{k}\bm{V}_{g_k}^{(t)}\!$ and
		$\vartheta_{{\rm m},k}^{(t)}=-{\rm Tr}\left(({\bm I}_{d_{\rm a}}-\bm{\varTheta}_{{\rm m},k}^{(t){\rm H}}{\bm H}_{k}\bm{V}_{g_k}^{(t)})^{-1}({\bm I}_{d_{\rm a}}+\delta_{\rm a}^{2}\bm{\varTheta}_{{\rm m},k}^{(t){\rm H}}\bm{\varTheta}_{{\rm m},k}^{(t)})\right)-\log_{2}\left|{\bm I}_{d_{\rm a}}-\bm{\varTheta}_{{\rm m},k}^{(t){\rm H}}\bm{H}_{k}\bm{V}_{g_k}^{(t)})\right|+d_{\rm a}$. 
	Moreover, $\bar{R}_{{\rm B},k}^{(t)}({\cal V})$ and $\bar{R}_{{\rm M},k}^{(t)}({\cal V})$ satisfy  two properties:
		\begin{itemize}
			\item[1)] $\bar{R}_{{\rm B},k}^{(t)}({\cal V})\leq R_{{\rm B},k}({\cal V})$ and $\bar{R}_{{\rm M},k}^{(t)}({\cal V})\leq R_{{\rm M},k}({\cal V})$, with the equalities holding at ${\cal V}={\cal V}^{(t)}$;
			\item[2)] $\nabla\bar{R}_{{\rm B},k}^{(t)}({\cal V}^{(t)}) = \nabla R_{{\rm B},k}({\cal V}^{(t)})$ and $\nabla\bar{R}_{{\rm M},k}^{(t)}({\cal V}^{(t)})= \nabla R_{{\rm M},k}({\cal V}^{(t)})$.
		\end{itemize}
	\end{prop}
	
	\begin{IEEEproof}
		Please refer to Appendix \ref{Appendix-A}.
	\end{IEEEproof}
	
	With \eqref{rate_mm} and \eqref{rate_uu}, the nonconvex \eqref{R_ck} and \eqref{R_gk} can be respectively approximated as
	\begin{align}
		{\cal R}_{0}-\bar{R}_{{\rm B}, k}^{(t)}(\mathcal{V}) & \leq 0, \ \forall k \in \mathcal{K}, \label{C22} \\
		{\cal R}_{g_{k}}-\bar{R}_{{\rm M},k}^{(t)}(\mathcal{V}) & \leq 0, \ \forall k \in \mathcal{K}. \label{C55}
	\end{align}
	Since $\bar{R}_{{\rm B},k}^{(t)}(\mathcal{V})$ and $\bar{R}_{{\rm M},k}^{(t)}(\mathcal{V})$ in \eqref{rate_mm} and \eqref{rate_uu} are concave quadratic over $\cal V$, the constraints \eqref{C22} and \eqref{C55} are convex.
	
	Now, we deal with the nonsmoothness of the indicator function $\mathds{1}\{x\}$ in \eqref{C1}. By using a similar approach as in \cite{rinaldi2010concave}, the nonsmooth $\mathds{1}\{x\}$ can be approximated by a smooth and convex $\ell_{1}$-norm, that is,
	\begin{equation}\label{l0_l1}
		\mathds{1}\left\{\left\|\bm{A}_{n}\bm{V}_{j}\right\|_{F}^{2}\right\}\approx \hat{\rho}_{{\rm v},n,j}^{(t)}\left\|\bm{A}_{n}\bm{V}_{j}\right\|_{F}^{2},
	\end{equation}
	where $\hat{\rho}_{{\rm v},n,j}^{(t)} \triangleq 1/(\|\bm{A}_{n}{\bm V}_{j}^{(t)}\|_{F}^{2}+\epsilon)$ is a weight factor with $\epsilon$ being a positive value to control the smoothness of the approximation and $\bm{V}_{j}^{(t)}$ being the solution obtained in the $t^{\rm th}$ iteration. By using the approximations \eqref{R_ck}-\eqref{R_gk} and \eqref{l0_l1}, the constraint in \eqref{C1} can be re-expressed as
	\begin{equation} \label{C_fhh}
		\sum_{j\in 0\cup{\cal G}}\hat{\rho}_{{\rm v},n,j}^{(t)}\left\|\bm{A}_{n}\bm{V}_{j}\right\|_{F}^{2}{\cal R}_{j}-R_{{\rm F},n}(\mathcal{V})\leq 0,\ \forall n\in{\cal N},
	\end{equation}
	which is still nonconvex because of the coupling between ${\cal R}_{j}$ and $\|\bm{A}_{n}\bm{V}_{j}\|_{F}^{2}$. 
	To address this issue, we replace ${\cal R}_{j}$ with $\hat{\cal R}_{j}$, that is, the value obtained in the previous iteration \cite{7942111}, then,  \eqref{C_fhh} can be simplified as
	\begin{equation} \label{C_ffh222}
		\hat{R}_{{\rm A},n}^{(t)}({\cal V})-R_{{\rm F},n}({\cal V})\leq 0,\ \forall n\in{\cal N},
	\end{equation}
	where $\hat{R}_{{\rm A},n}^{(t)}({\cal V})\triangleq \sum_{j\in 0\cup{\cal G}} \hat{\rho}_{{\rm v},n,j}^{(t)}\|\bm{A}_{n}\bm{V}_{j}\|_{F}^{2}\hat{\cal R}_{j}$. Due to the nonconvexity of $R_{{\rm F}, n}({\cal V})$, the constraint \eqref{C_ffh222} is nonconvex. It is noteworthy that the decoupling between the variables ${\cal R}_{j}$ and $\|\bm{A}_{n}\bm{V}_{j}\|_{F}^{2}$ is essentially an alternating optimization based on the popular block coordinate descent algorithm, which is proved to converge to the Karush-Kuhn-Tucker (KKT) point of the original optimization problem \cite[Prop. 3.7.1]{Nonlinear}. As a result, this decoupling does not only simplify the constraint, it also guarantees that \eqref{C_ffh222} is still a valid restriction of \eqref{C_fhh}.
		
	By using a similar approach as above, we can construct a sequence of concave quadratic functions to approximate $R_{{\rm F},n}(\mathcal{V})$ in \eqref{C1}, given below.
	
	\begin{prop} \label{Proposition-2}
		Given a fixed point $\mathcal{V}^{(t)}$, the lower bounding concave quadratic function for $R_{{\rm F},n}(\mathcal{V})$ is defined as
		\begin{equation} \label{rate_ff}
			\bar{R}_{{\rm F},n}^{(t)}({\cal V})\triangleq\vartheta_{{\rm f},n}^{(t)}-\sum_{\ell\in{\cal N}}{\rm Tr}({\bm U}_{\ell}^{\rm H}{\bm \varXi}_{{\rm f},n}^{(t)}{\bm U}_{\ell})-{\Re}\{{\rm Tr}({\bm \varUpsilon}_{{\rm f},n}^{(t)}{\bm U}_{n})\},
		\end{equation}
		where 
		${\bm{\varXi}_{{\rm f},n}^{(t)}=\bm{G}_{n}^{\rm H}\bm{\varTheta}_{{\rm f},n}^{(t)}({\bm I}_{d_{f}}-\bm{\varTheta}_{{\rm f},n}^{(t){\rm H}}
		\bm{G}_{n}\bm{U}_{n}^{(t)})^{-1}\bm{\varTheta}_{{\rm f},n}^{(t){\rm H}}\bm{G}_{n}}$, 
		$\bm{\varUpsilon}_{{\rm f},n}^{(t)}=-2({\bm I}_{d_{\rm f}}-\bm{\varTheta}_{{\rm f},n}^{(t){\rm H}}\bm{G}_{n}\bm{U}_{n}^{(t)})^{-1}{\bm \varTheta}_{{\rm f},n}^{(t){\rm H}}\bm{G}_{n}$,
		$\bm{\varTheta}_{{\rm f},n}^{(t)}=\left(\sum_{\ell\in{\cal N}}\bm{G}_{n}\bm{U}_{\ell}^{(t)}\bm{U}_{\ell}^{(t)\rm H}\bm{G}_{n}^{{\rm H}}+\sigma_{\rm f}^{2}\bm{I}_{M}\right)^{-1}\bm{G}_{n}\bm{U}_{n}^{(t)}$ and
		$\vartheta_{{\rm f},n}^{(t)}=-{\rm Tr}\left(({\bm I}_{d_{\rm f}}-\bm{\varTheta}_{{\rm f},n}^{(t){\rm H}}{\bm G}_{n}\bm{U}_{n}^{(t)})^{-1}({\bm I}_{d_{\rm f}}+\sigma_{\rm f}^{2}\bm{\varTheta}_{{\rm f},n}^{(t){\rm H}}\bm{\varTheta}_{{\rm f},n}^{(t)})\right)-\log_2\left|{\bm I}_{d_{\rm f}}-\bm{\varTheta}_{{\rm f},n}^{(t){\rm H}}\bm{G}_{n}\bm{U}_{n}^{(t)}\right|+d_{\rm f}$.
		Also, $\bar{R}_{{\rm f},n}^{(t)}({\cal V})$ satisfies two properties: 
		\begin{itemize}
			\item[1)] $\bar{R}_{{\rm F},n}^{(t)}({\cal V})\leq R_{{\rm F},n}({\cal V})$, with the equality holding at ${\cal V}={\cal V}^{(t)}$;
			\item[2)] $\nabla\bar{R}_{{\rm F},n}^{(t)}({\cal V}^{(t)})= \nabla R_{{\rm F},n}({\cal V}^{(t)})$.
		\end{itemize}	
	\end{prop}
	\begin{IEEEproof}
		The proof is similar to that in Appendix \ref{Appendix-A} and thus omitted for brevity.
	\end{IEEEproof}
	
	By virtue of Proposition~\ref{Proposition-2} and with a fixed point  $\mathcal{V}^{(t)}$, the nonconvex constraint \eqref{C_ffh222} can be approximated as 
	\begin{equation} \label{C_ffh23}
		\sum_{j\in 0\cup{\cal G}}\hat{\rho}_{{\rm v},n,j}^{(t)}\left\|\bm{A}_{n}\bm{V}_{j}\right\|_{F}^{2}\hat{\cal R}_{j}-\bar{R}_{{\rm F},n}^{(t)}({\cal V})\leq 0,\ \forall n\in{\cal N},
	\end{equation}
	which is convex, as $\|\bm{A}_{n}\bm{V}_{j}\|_{F}^{2}$ and $\bar{R}_{{\rm F},n}^{(t)}({\cal V})$ are convex and concave quadratic over ${\cal V}$, respectively.
	
	Finally, we tackle the nonconvex constraint \eqref{C2}, which can be re-expressed as
	\begin{equation} \label{EHC}
		\mathcal{F}^{-1}\left(e_{q}\right)-P_{{\rm R}, q}({\cal V}) \leq 0,\ \forall q \in \mathcal{Q},
	\end{equation}
	where
	\begin{align} \label{neh}
	\!\mathcal{F}^{-1}\left(x\right) &= \begin{cases}
	\begin{array}{ll}
	+\infty, & \text{if } x \geq P_{\max} \\
	\frac{\iota_{2}}{\iota_{1}}-\frac{1}{\iota_{1}}\ln\left(\frac{1+\eth_{1}}{1+\eth_{2}x}-1\right), & \text{if } 0<x<P_{\max} \\
	0, & \text{if } x \leq 0
	\end{array} \end{cases}\!
	\end{align}
	denotes the pseudo-inverse of $\mathcal{F}\left(x\right)$, with  $\eth_{1} \triangleq \exp\left(-\iota_{1} P_{0}+\iota_{2}\right)$ and $\eth_{2} \triangleq P_{\max}^{-1}\exp\left(-\iota_{1} P_{0}+\iota_{2}\right)$ \cite{8322450}. In view of $P_{{\rm R}, q}({\cal V})$ given by \eqref{RF_i}, it is clear that ${\rm Tr}\left(\bm{V}_{j}^{\rm H}\bm{F}_{q}^{\rm H}\bm{F}_{q}\bm{V}_{j}\right)$ is a convex function of ${\bm V}_{j}$, which can be approximated by its first-order Taylor series expansion:
	\begin{align}\label{FOT}
		{\rm Tr}\left(\bm{V}_{j}^{\rm H}\bm{F}_{q}^{\rm H}\bm{F}_{q}\bm{V}_{j}\right)\geq &-{\rm Tr}\left(\bm{V}_{j}^{(t)\rm H}\bm{F}_{q}^{\rm H}\bm{F}_{q}\bm{V}_{j}^{(t)}\right)\nonumber\\
		&{}+2\Re\left\{{\rm Tr}\left(\bm{V}_{j}^{(t)\rm H}\bm{F}_{q}^{\rm H}\bm{F}_{q}\bm{V}_{j}\right)\right\}.
	\end{align}
	Then, an approximation of $P_{{\rm R},q}(\mathcal{V})$ can be expressed as
	\begin{equation} \label{phi_ee1}
		\!P_{{\rm R},q}({\cal V}) \geq \varphi_{{\rm e},q}^{(t)}+2\sum_{j\in0\cup{\cal G}}{\Re}\left\{{\rm Tr}\big(\bm{V}_{j}^{(t)\rm H}\bm{F}_{q}^{\rm H}
		\bm{F}_{q}\bm{V}_{j}\big)\right\}\triangleq\bar{P}_{{\rm R}, q}^{(t)}({\cal V}), \!
	\end{equation}
	where $\varphi_{{\rm e},q}^{(t)} \triangleq -\sum_{j\in 0\cup{\cal G}}{\rm Tr}\left(\bm{V}_{j}^{(t)\rm H}\bm{F}_{q}^{\rm H}\bm{F}_{q}\bm{V}_{j}^{(t)}\right)$. As a result, with a fixed point $\mathcal{V}^{(t)}$, the nonconvex constraint \eqref{C2} can be approximated as
	\begin{equation} \label{C44}
		{\cal F}^{-1}(e_{q})- \bar{P}_{{\rm R}, q}^{(t)}({\cal V}) \leq 0, \ \forall q \in {\cal Q},
	\end{equation}
	which is convex as $\bar{P}_{{\rm R}, q}^{(t)}({\cal V})$ is linear over ${\cal V}$.
	
	Now, with the convex constraints obtained in \eqref{C22}-\eqref{C55}, \eqref{C_ffh23} and \eqref{C44}, problem \eqref{P0} can be iteratively solved in the SCA framework, with the $t^{\rm th}$ SCA subproblem explicitly given by
	\begin{subequations} \label{P_t1}
		\begin{align}
			\{{\cal V}^{(t+1)},{\cal R}^{(t+1)}\}&=\min_{\mathcal{V}, \mathcal{R}}\ -\sum_{j\in0\cup{\cal G}}\alpha_{j}{\cal R}_{j}, \label{objj} \\
			{\rm s.t.}
			& \ \eqref{C3},\eqref{C4}, \eqref{C22},\eqref{C55}, \eqref{C_ffh23}, \eqref{C44}. \label{tgg}
		\end{align}
	\end{subequations}
	Clearly, the subproblem \eqref{P_t1} is convex as the objective function \eqref{objj} is linear and all the constraints are convex. In the next subsection,  a low-complexity first-order algorithm is developed, instead of traditional second-order interior-point methods with high complexity.
	
	\subsection{First-Order Algorithm in Dual Domain}
	Unlike traditional second-order algorithms, first-order approachs that alternatively perform a gradient step and a projection step \cite{Beck2017}, need only gradient information and thus enjoy much lower computational complexity, making them more suitable for handling large-scale optimization problems. However, as problem \eqref{P_t1} is imposed by coupling constraints, the projection onto them would be highly complicated, if not properly handled.
	
	To address this issue, we transform \eqref{P_t1} by majorizing the cost function \eqref{objj} with a strongly convex upper bound. In particular, given any fixed point ${\cal V}^{(t)}$, \eqref{objj} can be strongly convexified by adding two positive quadratic terms:
	\begin{align}\label{obj_up}
		\Gamma^{(t)}({\cal V,R}) &=-\sum_{j\in 0\cup{\cal G}}\alpha_{j}{\cal R}_{j}+\rho_{1}\sum_{j\in0\cup{\cal G}}\|\bm{V}_{j}-\bm{V}_{j}^{(t)}\|_{F}^{2}\nonumber\\
		&\quad {}+\rho_{2}\sum_{n\in{\cal N}}\|\bm{U}_{n}-\bm{U}_{n}^{(t)}\|^{2}_{F},
	\end{align}
	where $\rho_{1}$ and $\rho_{2}$ are fixed positive parameters. In principle, the added \textit{proximal term} $\rho_{1}\sum_{j\in0\cup{\cal G}}\|\bm{V}_{j}-\bm{V}_{j}^{(t)}\|_{F}^{2}+\rho_{2}\sum_{n\in{\cal N}}\|\bm{U}_{n}-\bm{U}_{n}^{(t)}\|_{F}^{2}$ is to make the objective function strongly concave with respect to ${\cal V}$. According to \cite[Prop. 4.1]{bertsekas1989parallel}, the iterations $\{{\cal V}^{(t)}\}$ is guaranteed to converge to a limit point. Consequently, \eqref{obj_up} serves as a tight upper bound of \eqref{objj}, with their function values being equal at ${\cal V}={\cal V}^{(t)}$. Therefore, \eqref{P_t1} can be transformed into
	\begin{equation} \label{P_0_up}
		\{{\cal V}^{(t+1)},{\cal R}^{(t+1)}\}=\min_{{\cal R}, {\cal V}}\ \Gamma^{(t)}({\cal V,R}), \text{ s.t. } \eqref{tgg}.
	\end{equation}
	With the strong convexity of $\Gamma^{(t)}({\cal V,R})$, we can derive the dual problem of \eqref{P_0_up}. Specifically, let ${\cal L}\triangleq\left\{\left\{\lambda_{{\rm r},n},\lambda_{{\rm f},n}\right\}_{n\in{\cal N}}, \lambda_{\rm c}, \left\{\lambda_{{\rm b},k}, \lambda_{{\rm m},k}\right\}_{k\in{\cal K}}, \left\{\lambda_{{\rm e},q}\right\} _{q\in{\cal Q}}\right\}$ be the dual variables corresponding to the constraints shown in \eqref{tgg}, the dual problem of \eqref{P_0_up} is formalized below in closed-form.
	\begin{prop}
		\label{Prop_2}
		\begin{subequations}
			The dual problem of \eqref{P_0_up} can be explicitly expressed as
			\begin{align} \label{dual}
				\lefteqn{\max_{{\cal L}\in{\cal P}}\ {\cal D}\left({\cal L}\right)} \nonumber \\
				& = \rho_{1}\sum_{j\in0\cup{\cal G}}\|\bm{V}_{j}^{\lozenge}-\bm{V}_{j}^{(t)}\|_{F}^{2}+\rho_{2}\sum_{n\in{\cal N}}\|\bm{U}_{n}^{\lozenge}-\bm{U}_{n}^{(t)}\|_{F}^{2}\nonumber\\
				& \quad {}+\lambda_{\rm c}\Big(\sum_{n\in{\cal N}}\left\|{\bm U}_{n}^{\lozenge}\right\|_{F}^{2}-p_{c}\Big)+\sum_{q\in{\cal Q}}\lambda_{{\rm e},q}\big({\cal F}^{-1}\left(e_{q}\right)\nonumber\\
				& \quad {}-\bar{P}_{{\rm R},q}^{(t)}({\cal V}^{\lozenge})\big)-\sum_{k\in{\cal K}}\Big[\lambda_{{\rm m},k}\bar{R}_{{\rm M},k}^{(t)}({\cal V}^{\lozenge})+\lambda_{{\rm b},k}\nonumber\\
				& \quad {}\times\bar{R}_{{\rm B},k}^{(t)}({\cal V}^{\lozenge})\Big]
				  +\sum_{n\in{\cal N}}\Big[\lambda_{{\rm f},n}\left(\hat{R}_{{\rm R},n}({\cal V}^{\lozenge})-\bar{R}_{{\rm F},n}^{(t)}({\cal V}^{\lozenge})\right)\nonumber\\
				& \quad {}+\lambda_{{\rm r},n}\left(P_{{\rm T},n}({\cal V}^{\lozenge})-p_{n}\right)\Big],
			\end{align}
			where $\mathcal{V}^{\lozenge} \triangleq \left\{ \left\{ {\bm V}_{j}^{\lozenge}\right\}_{j \in 0\cup{\mathcal G}},  \left\{{\bm U}_{n}^{\lozenge}\right\}_{n \in \mathcal{N}}\right\}$ is uniquely given by
			\begin{align}
				{\bm V}_{j}^{\lozenge} &= \left(\mathcal{\bm A}_{j}^{(t)}\right)^{-1}\bm{A}_{j}^{(t)}, \ \forall j\in0\cup{\cal G},\label{dp_m1} \\
				{\bm U}_{n}^{\lozenge} &=\left(\mathcal{\bm B}_{n}^{(t)}\right)^{-1}\bm{B}_{n}^{(t)}, \ \forall n\in{\cal N},\label{dp_m3}
			\end{align}
		with $\mathcal{\bm A}_{j}^{(t)}$ and ${\bm A}_{j}^{(t)}$ are given by \eqref{a1} and \eqref{a2}, respectively, shown in the top of the next page, and
\begin{figure*}[tbh]
	\begin{align}
		\mathcal{\bm A}_{j}^{(t)}& \triangleq 
		\begin{cases}
		\begin{array}{ll}
		\sum_{n\in{\cal N}}\bm{\Omega}_{n,j}+\rho_{1}\bm{I}_{MN}+\sum_{k\in{\cal K}}\lambda_{{\rm b},k}\bm{\varXi}_{{\rm b},k}^{(t)}, & {\rm if}\ j=0,\\
		\sum_{n\in{\cal N}}\bm{\Omega}_{n,j}+\rho_{1}\bm{I}_{MN}+\sum_{k\in{\cal K}}(\lambda_{{\rm b},k}\bm{\varXi}_{{\rm b},k}^{(t)}
		+\lambda_{{\rm m},k}\bm{\varXi}_{{\rm m},k}^{(t)}), & {\rm if}\ j\in{\cal G},
		\end{array}
		\end{cases}\label{a1}\\
		\bm{A}_{j}^{(t)}& \triangleq 
		\begin{cases}
		\begin{array}{ll}
		\rho_{1}\bm{V}_{0}^{(t)}+\sum_{q\in{\cal Q}}\lambda_{{\rm e},q}\bm{F}_{q}^{\rm H}\bm{F}_{q}\bm{V}_{0}^{(t)}-\frac{1}{2}\sum_{k\in{\cal K}}\lambda_{{\rm b},k}\bm{\varUpsilon}_{{\rm b},k}^{(t)},  & {\rm if}\ j=0,\\
		\rho_{1}\bm{V}_{g}^{(t)}+\sum_{q\in{\cal Q}}\lambda_{{\rm e},q}\bm{F}_{q}^{\rm H}\bm{F}_{q}\bm{V}_{g}^{(t)}-\frac{1}{2}\sum_{k\in{\cal K}_{g}}\lambda_{{\rm m},k}\bm{\varUpsilon}_{{\rm m},k}^{(t)},  & {\rm if}\ j\in{\cal G},
		\end{array}
		\end{cases}\label{a2}
	\end{align}
	\rule[0.5ex]{2\columnwidth}{0.5pt}
\end{figure*}
	\begin{align}
		\mathcal{\bm B}_{n}^{(t)}&\triangleq\left(\lambda_{\mathrm{c}}+\rho_{2}\right)\bm{I}_{L}+\sum_{\ell\in{\cal N}}\lambda_{\mathrm{f}, \ell} \bm{\varXi}_{{\rm f},\ell}^{(t)},\\
		{\bm B}_{n}^{(t)}&\triangleq \rho_{2}\bm{U}_{n}^{(t)}-\frac{\lambda_{\mathrm{f},n}}{2} \bm{\varUpsilon}_{{\rm f},n}^{(t){\rm H}}.\label{bb}		
	\end{align}
			where $\bm{\Omega}_{n,j} \triangleq {\rm diag}\left\{\left[{\bm 0}_{(n-1)M},\varkappa_{{\rm v},n,j}^{(t)\frac{1}{2}}{\bm 1}_{M},{\bm 0}_{(N-n)M}\right]\right\}$ with $\varkappa_{{\rm v},n,j}^{(t)} \triangleq \lambda_{{\rm f},n}\hat{\rho}_{{\rm v},n,j}^{(t)}\hat{\cal R}_{j}+\lambda_{{\rm r},n}$. Moreover, the domain of the dual function $\mathcal{D}\left(\mathcal{L}\right)$ is 
			\begin{align} \label{dom}
				\mathcal{P} &= 
				\left\{
				\begin{array}{rl}
					\lambda_{{\rm m}, k},\ \lambda_{{\rm c}, k} \geq 0, & \forall k \in \mathcal{K}; \\
					\lambda_{{\rm f}, n},\ \lambda_{{\rm r}, n} \geq 0, & \forall n \in \mathcal{N}; \\
					\lambda_{\rm c} \geq 0,\ \lambda_{{\rm e}, q} \geq 0, & \forall q \in \mathcal{Q}; \\
					\sum_{k \in \mathcal{K}_{g}}\limits\lambda_{{\rm m}, k}-\alpha_{g}=0, & \forall g \in \mathcal{G};\\
					\sum_{k \in \mathcal{K}}\limits\lambda_{{\rm b}, k}-\alpha_{0}=0. &
				\end{array}
				\right.
			\end{align}
		\end{subequations}
	\end{prop}
	
	\begin{IEEEproof}
		See Appendix~\ref{Appendix-B}.
	\end{IEEEproof}
	
	In view of Proposition~\ref{Prop_2}, for any fixed $\mathcal{L}$, the value of $\mathcal{V}^{\lozenge}$ is uniquely determined by \eqref{dp_m1}-\eqref{dp_m3}. Also, it is straightforward to compute the partial derivatives as
	$\frac{\partial \mathcal{D}(\mathcal{L})}{\partial \lambda_{{\rm m},k}} = - \bar{R}_{{\rm M},k}^{(t)}(\mathcal{V}^{\lozenge})$,
	$\frac{\partial \mathcal{D}(\mathcal{L})}{\partial \lambda_{{\rm b},k}} = -\bar{R}_{{\rm B},k}^{(t)}(\mathcal{V}^{\lozenge}), \forall k\in{\cal K}$;
	$\frac{\partial \mathcal{D}(\mathcal{L})}{\partial \lambda_{{\rm e},q}} = \mathcal{F}^{-1}\left(e_{q}\right)
	-\bar{P}_{{\rm R}, q}^{(t)}(\mathcal{V}^{\lozenge})$, $\forall q \in \mathcal{Q}$; 
	$\frac{\partial \mathcal{D}(\mathcal{L})}{\partial \lambda_{{\rm f},n}} = \hat{R}_{{\rm A},n}^{(t)}({\cal V}^{\lozenge})-\bar{R}_{{\rm F},n}^{(t)}({\cal V}^{\lozenge})$,
	$\frac{\partial \mathcal{D}(\mathcal{L})}{\partial \lambda_{{\rm r},n}} = P_{{\rm T},n}(\mathcal{V}^{\lozenge})-p_{n}$, $\forall n \in \mathcal{N}$, and $\frac{\partial \mathcal{D}(\mathcal{L})}{\partial \lambda_{{\rm c}}} = \sum_{n \in \mathcal{N}}\left\|\bm{U}_{n}^{\lozenge}\right\|_{F}^{2}-p_{c}$.
	Consequently, the gradient update step at the $s^{\rm th}$ iteration $\mathcal{D}(\left\{\mu_{{\rm m}, k}, \mu_{{\rm b}, k}\right\}_{k \in \mathcal{K}},  \left\{\mu_{{\rm e}, q}\right\}_{q\in{\cal Q}}, \left\{\mu_{{\rm f},n}, \mu_{{\rm p}, n}\right\}_{n \in \mathcal{N}}, \mu_{\rm c})$ can be expressed as
	\begin{align} \label{La_fac}
		\left\{
		\begin{array}{ll}
			\mu_{{\rm m}, k}^{(s)} \leftarrow \lambda_{{\rm m}, k}^{(s)} - \nu_{s}\bar{R}_{{\rm M},k}^{(t)}(\mathcal{V}^{\lozenge}), & \forall k \in \mathcal{K},\\
			\mu_{{\rm b}, k}^{(s)} \leftarrow \lambda_{{\rm b}, k}^{(s)} -\nu_{s}\bar{R}_{{\rm B}, k}^{(t)}(\mathcal{V}^{\lozenge}), & \forall k \in \mathcal{K},\\
			\mu_{{\rm e}, q}^{(s)} \leftarrow \lambda_{{\rm e}, q}^{(s)} + \nu_{s}\left(\mathcal{F}^{-1}(e_{q}) - \bar{P}_{{\rm R}, q}^{(t)}(\mathcal{V}^{\lozenge})\right), & \forall q \in \mathcal{Q}, \\
			\mu_{{\rm f}, n}^{(s)} \leftarrow \lambda_{{\rm f}, n}^{(s)} + \nu_{s}\left( \hat{R}_{{\rm A},n}^{(t)}({\cal V}^{\lozenge})-\bar{R}_{{\rm F},n}^{(t)}({\cal V}^{\lozenge})\right), & \forall n \in \mathcal{N}, \\
			\mu_{{\rm r}, n}^{(s)} \leftarrow \lambda_{{\rm r}, n}^{(s)} + \nu_{s}\left(P_{{\rm T}, n}(\mathcal{V}^{\lozenge}) - p_{n}\right), & \forall n \in \mathcal{N},\\
			\mu_{\rm c}^{(s)} \leftarrow \lambda_{\rm c}^{(s)} + \nu_{s}\Big(\sum_{n \in \mathcal{N}}\limits\left\|\bm{U}_{n}^{\lozenge}\right\|_{F}^{2}-p_{c}\Big), & 
		\end{array}
		\right.
	\end{align}
	where $\nu_{s}$ denotes the step size at the $s^{\rm th}$ iteration. To satisfy the constraints in \eqref{dom}, we need to further project
	$\!\left\{ \{\mu_{{\rm m}, k}^{(s)}, \mu_{{\rm b}, k}^{(s)}\}_{k \in \mathcal{K}}, \{\mu_{{\rm e}, q}^{(s)}\}_{q\in{\cal Q}}, \{\mu_{{\rm f},n}^{(s)}, \mu_{{\rm r}, n}^{(s)}\}_{n \in \mathcal{N}}, \mu_{\rm c}^{(s)}\right\}\!$ onto $\mathcal{P}$ to find its nearest feasible point, which is equivalent to
	\begin{align}
		\min_{\mathcal{L} \in \mathcal{P}}\ 
		& \sum_{n \in \mathcal{N}}\left[\left(\lambda_{{\rm f}, n}^{(s+1)}-\mu_{{\rm f}, n}^{(s)}\right)^{2} 
		+ \left(\lambda_{{\rm r}, n}^{(s+1)}-\mu_{{\rm r}, n}^{(s)}\right)^{2}\right]\nonumber\\
		&{}+ \sum_{k \in \mathcal{K}}\left[\left(\lambda_{{\rm m}, k}^{(s+1)}-\mu_{{\rm m}, k}^{(s)}\right)^{2} 
		+ \left(\lambda_{{\rm b}, k}^{(s+1)}-\mu_{{\rm b}, k}^{(s)}\right)^{2}\right]\nonumber\\
		&{}+\left(\lambda_{\rm c}^{(s+1)}-\mu_{\rm c}^{(s)}\right)^{2}+\sum_{q\in{\cal Q}}\left(\lambda_{{\rm e}, q}^{(s+1)}-\mu_{{\rm e}, q}^{(s)}\right)^{2}. \label{pro_1}
	\end{align}
	Since $\mathcal{P}$ in \eqref{dom} consists of separable linear constraints, we can easily find the projection result of \eqref{pro_1} as
	(for more details, please refer to  Appendix~\ref{Appendix-C}):
	\begin{align}
		\!\lambda_{{\rm m},k}^{(s+1)} &= \left(\mu_{{\rm m},k}^{(s)}-\frac{\varpi_{g_{k}}}{2}\right)^{+}, \label{pro_22} \\  
		\lambda_{{\rm b},k}^{(s+1)}&= \left(\mu_{{\rm b},k}^{(s)}-\frac{\varpi_{0}}{2}\right)^{+},\ \forall k\in{\cal K}; \label{pro_23} \\
		\lambda_{{\rm f},n}^{(s+1)} &=\left(\mu_{{\rm f},n}^{(s)}\right)^{+}, \ 
		\lambda_{{\rm r},n}^{(s+1)} =\left(\mu_{{\rm r},n}^{(s)}\right)^{+}, \ \forall n\in{\cal N}; \label{pro_23-a} \\
		\lambda_{{\rm e},q}^{(s+1)}& = \left(\mu_{{\rm e}, q}^{(s)}\right)^{+}, \ \forall q\in{\cal Q};\ \lambda_{\rm c}^{(s+1)} =(\mu_{\rm c}^{(s)})^{+}, \label{pro_2}
	\end{align}
	where $(\cdot)^{+}$ means the non-negative projection on \eqref{dom}, and $\varpi_{g_{k}}$ and $\varpi_{0}$ are parameters satisfying
	\begin{align}
		\sum_{k \in \mathcal{K}_{g}}\left(\mu_{{\rm m}, k}^{(s)}-\frac{\varpi_{g_{k}}}{2}\right)^{+}&=\alpha_{g},\ \forall g\in{\cal G},\\
		\sum_{k \in \mathcal{K}}\left(\mu_{{\rm b}, k}^{(s)}-\frac{\varpi_{0}}{2}\right)^{+}&=\alpha_{0},
	\end{align}
	whose values can be readily determined by using the bisection method.
	
	By iteratively updating $\mathcal{L}$ as per \eqref{La_fac} and \eqref{pro_22}-\eqref{pro_2}, we can get the optimal solution of $\mathcal{L}$  to the dual problem \eqref{dual}. Then, the optimal solution of $\mathcal{V}$ to the primal problem \eqref{P_0_up} is obtained by substituting the optimal $\mathcal{L}$ into \eqref{dp_m1}-\eqref{dp_m3}. In light of \eqref{cpp_1}, the optimal solutions of $\left\{{\cal R}_{0},\left\{{\cal R}_{g}\right\}_{g\in {\cal G}}\right\}$ to \eqref{P_0_up} satisfies its equality, that is, 
	\begin{equation}
		\mathcal{R}_{0} = \min_{k \in {\cal K}}\{\bar{R}_{{\rm B},k}^{(t)}({\cal V})\},\ 
		\mathcal{R}_{g} = \min_{k \in \mathcal{K}_{g}}\{\bar{R}_{{\rm M},k}^{(t)}({\cal V})\}.\label{R_ggg}
	\end{equation}
	
	\begin{algorithm}[!t]
		\small
		\caption{First-Order Algorithm for Solving Problem \eqref{P0}}
		\label{Algorithm1}
		\begin{algorithmic}[1]
			\STATE \textbf{Initialization:} Generate $\mathcal{V}^{(0)}$ via Algorithm~\ref{Algorithm2} (to be detailed in Section IV).
			\STATE {$t = 0$;}
			\REPEAT
			\STATE {Compute $\bm{\varXi}_{{\rm b}, k}^{(t)}$, 
				$\bm{\varXi}_{{\rm f}, n}^{(t)}$
				$\bm{\varXi}_{{\rm m}, k}^{(t)}$, 	
				$\bm{\varUpsilon}_{{\rm b}, k}^{(t)}$, 	
				$\bm{\varUpsilon}_{{\rm f}, n}^{(t)}$,
				$\bm{\varUpsilon}_{{\rm m}, k}^{(t)}$, 	
				and
				$\varphi_{{\rm e}, q}^{(t)}$, $\forall q \in \mathcal{Q}$,
				$\hat{\rho}_{{\rm v},n,j}^{(t)}$,			
				$\hat{\cal R}_{j}$, $\forall j \in 0\cup{\cal G}$,
				$\forall n \in \mathcal{N}$ according to \eqref{rate_ff}-\eqref{rate_uu} and \eqref{phi_ee1};}
			\STATE {$\mathcal{V} = \mathcal{V}^{(t)}$};
			\STATE {Set $\lambda_{\rm c}=\lambda_{{\rm m}, k} =
				\lambda_{{\rm b}, k} = \lambda_{{\rm e}, q} = \lambda_{{\rm r}, n} = \lambda_{{\rm f}, n} = 0$, $\forall n \in \mathcal{N}, k \in \mathcal{K}, q \in \mathcal{Q}$;}
			\REPEAT
			\STATE{Update $\mathcal{V}^{\lozenge}$ as per \eqref{dp_m1}-\eqref{dp_m3};}
			\STATE{Update $\mathcal{L}$ according to \eqref{La_fac} and \eqref{pro_22}-\eqref{pro_2};}
			\UNTIL {convergence}
			\STATE {Output $\mathcal{V} = \mathcal{V}^{\lozenge}$;}
			\STATE {Output $\mathcal{V}^{(t+1)} = \mathcal{V}$;}
			\STATE {$t \leftarrow t+1$;}
			\UNTIL {convergence}
		\end{algorithmic}
	\end{algorithm}
	
    To sum up, we formalize the main procedure for finding the solution of problem \eqref{P0} through two-layer iterations in Algorithm~\ref{Algorithm1}. 
	For the outer iteration, \eqref{P0} is iteratively solved in the SCA framework over $t$, while for the inner iteration, each SCA subproblem \eqref{P_0_up} is iteratively solved over $s$. Since the domain of $\mathcal{D}(\mathcal{L})$ in \eqref{dom} is closed and convex, the iteration with respect to $s$ is guaranteed to converge to the global optimum of \eqref{dual} at a rate of $\mathcal{O}(1/s)$, if the step size $\nu_{s}$ is smaller than the inverse of the Lipschitz constant of $\nabla \mathcal{D}(\mathcal{L})$ \cite{beck2009gradient}. Moreover, since the primal problem \eqref{P_0_up} is convex, the convergent optimum of \eqref{dual} is also the global optimum of \eqref{P_0_up}, given that \eqref{P_0_up} is strictly feasible \cite{boyd_convex_2004}. It is noteworthy that, although \eqref{C_ffh222} and \eqref{obj_up} are exploited to improve the algorithm efficiency, Algorithm~1 is guaranteed to converge to a stationary point of problem \eqref{P0}\cite[Prop. 3.7.1]{Nonlinear}, \cite[Prop. 4.1]{bertsekas1989parallel}.
	
	It remains to mention that, since Algorithm~\ref{Algorithm1} is based on the SCA framework, it requires a feasible point of \eqref{P0} to start the SCA procedure, which is addressed in the next section.
	
	\begin{remark}[On the coordination overhead] 
			As channel estimation is performed locally at each RRH, there is no overhead for exchanging the instantaneous CSI among the RRHs. On the other hand, each iteration of Algorithm~\ref{Algorithm1} can be executed in parallel, which benefits lower coordination overhead. In particular, according to \eqref{dp_m1}-\eqref{bb}, the all primal variables $\mathcal{V}^{\lozenge} \triangleq \left\{ \left\{ {\bm{V}}_{j}^{\lozenge}\right\}_{j \in 0\cup\mathcal{G}}, \left\{ {\bm{U}}_{n}^{\lozenge}\right\}_{n \in \mathcal{N}}\right\}$ can be updated in parallel. Since the update of each primal variable only depends on a few dual variables, the message passing overhead is small. Likewise, according to \eqref{La_fac} and \eqref{pro_22}-\eqref{pro_2}, the dual variables can also be updated in parallel and each depends on only a few primal variables. Thanks to the parallel structure, Algorithm~\ref{Algorithm1} has the potential of leveraging the modern multicore multi-thread processor architecture for speeding up the computation, with extremely low coordination overhead.
	\end{remark}

	\subsection{Acceleration with Momentum Techniques}
	As Algorithm~\ref{Algorithm1} only involves the gradient information, it may spend a large number of iterations to converge. To improve the convergence speed, Nesterov gradient method is widely used to accelerate the global convergence by exploiting Nesterov momentums, see e.g., \cite{beck2009fast, 8976409, 9184916}.  However, Nesterov method has a slow local convergence rate. Fortunately, the classic heavy-ball method is capable of improving local convergence rate \cite{polyak1964some}. Most recently, by jointly exploiting Nesterov and heavy-ball momentums, the work \cite{9222312} developed a double accelerated algorithm, allowing rapid acceleration in both global and local convergence rates. To be specific, according to \cite{9222312}, the gradient update step \eqref{La_fac} in Algorithm~\ref{Algorithm1} is recalculated as per the two substeps, given by \eqref{La_fac_AA_1} shown at the top of the next page, 
	\begin{subequations}
			\begin{figure*}[tbh]
		\begin{equation} \label{La_fac_AA_1}
			\left\{
			\begin{array}{ll}
				\tilde{\mu}_{{\rm m}, k}^{(s)} \leftarrow \lambda_{{\rm m}, k}^{(s)} - \nu_{s}\bar{R}_{{\rm M},k}^{(t)}(\mathcal{V}^{\lozenge})+\frac{\pi^{(s-1)}-1}{\pi^{(s)}}
				\left(\lambda_{{\rm m}, k}^{(s)}-\lambda_{{\rm m}, k}^{(s-1)}\right), & \forall k \in \mathcal{K},\\
				\tilde{\mu}_{{\rm b}, k}^{(s)} \leftarrow \lambda_{{\rm b}, k}^{(s)} - \nu_{s}\bar{R}_{{\rm B}, k}^{(t)}(\mathcal{V}^{\lozenge})+\frac{\pi^{(s-1)}-1}{\pi^{(s)}}
				\left(\lambda_{{\rm b}, k}^{(s)}-\lambda_{{\rm b}, k}^{(s-1)}\right), & \forall k \in \mathcal{K},\\
				\tilde{\mu}_{{\rm e}, q}^{(s)} \leftarrow \lambda_{{\rm e}, q}^{(s)} + \nu_{s}\left(\mathcal{F}^{-1}\left(e_{q}\right) - \bar{P}_{{\rm R}, q}^{(t)}(\mathcal{V}^{\lozenge})\right)+\frac{\pi^{(s-1)}-1}{\pi^{(s)}}
				\left(\lambda_{{\rm e}, q}^{(s)}-\lambda_{{\rm e}, q}^{(s-1)}\right), & \forall q \in \mathcal{Q}, \\
				\tilde{\mu}_{{\rm f}, n}^{(s)} \leftarrow \lambda_{{\rm f}, n}^{(s)} + \nu_{s}\left(\hat{R}_{{\rm A},n}^{(t)}({\cal V}^{\lozenge})-\bar{R}_{{\rm F},n}^{(t)}({\cal V}^{\lozenge})\right)+\frac{\pi^{(s-1)}-1}{\pi^{(s)}}
				\left(\lambda_{{\rm f}, n}^{(s)}-\lambda_{{\rm f}, n}^{(s-1)}\right), & \forall n \in \mathcal{N}, \\
				\tilde{\mu}_{{\rm r}, n}^{(s)} \leftarrow \lambda_{{\rm r}, n}^{(s)} + \nu_{s}\left(P_{{\rm T}, n}(\mathcal{V}^{\lozenge}) - p_{n}\right)+\frac{\pi^{(s-1)}-1}{\pi^{(s)}}
				\left(\lambda_{{\rm r}, n}^{(s)}-\lambda_{{\rm r}, n}^{(s-1)}\right), & \forall n \in \mathcal{N},\\
				\tilde{\mu}_{\rm c}^{(s)} \leftarrow \lambda_{\rm c}^{(s)} + \nu_{s}\left(\sum_{n \in \mathcal{N}}\left\|\bm{U}_{n}^{\lozenge}\right\|_{F}^{2}-p_{c}\right)+\frac{\pi^{(s-1)}-1}{\pi^{(s)}}
				\left(\lambda_{\rm c}^{(s)}-\lambda_{\rm c}^{(s-1)}\right), & 
			\end{array}
			\right.
		\end{equation}
						\rule[0.5ex]{2\columnwidth}{0.5pt}
	\end{figure*}
	where $\lambda_{{\rm m}, k}^{(s)}-\lambda_{{\rm m}, k}^{(s-1)}$, 
		$\lambda_{{\rm b}, k}^{(s)}-\lambda_{{\rm b}, k}^{(s-1)}$,
		$\lambda_{{\rm e}, q}^{(s)}-\lambda_{{\rm e}, q}^{(s-1)}$, 
		$\lambda_{{\rm f}, n}^{(s)}-\lambda_{{\rm f}, n}^{(s-1)}$,
		$\lambda_{{\rm r}, n}^{(s)}-\lambda_{{\rm r}, n}^{(s-1)}$ and
		$\lambda_{\rm c}^{(s)}-\lambda_{\rm c}^{(s-1)}$ denote the heavy-ball momentums, 
		and
		\begin{equation} \label{La_fac_AA_2}
			 \left\{ 
			\begin{array}{rl}
				\mu_{{\rm m},k}^{(s)}\leftarrow\tilde{\mu}_{{\rm m}, k}^{(s)}+\frac{\pi^{(s-1)}-1}{\pi^{(s)}}
				\left(\tilde{\mu}_{{\rm m}, k}^{(s)}-\tilde{\mu}_{{\rm m}, k}^{(s-1)}\right), &  \forall k \in \mathcal{K}, \\
				\mu_{{\rm b},k}^{(s)}\leftarrow\tilde{\mu}_{{\rm b}, k}^{(s)}+\frac{\pi^{(s-1)}-1}{\pi^{(s)}}
				\left(\tilde{\mu}_{{\rm b}, k}^{(s)}-\tilde{\mu}_{{\rm b}, k}^{(s-1)}\right), &  \forall k \in \mathcal{K}, \\
				\mu_{{\rm e},q}^{(s)}\leftarrow\tilde{\mu}_{{\rm e}, q}^{(s)}+\frac{\pi^{(s-1)}-1}{\pi^{(s)}}
				\left(\tilde{\mu}_{{\rm e}, q}^{(s)}-\tilde{\mu}_{{\rm e}, q}^{(s-1)}\right), &  \forall q \in \mathcal{Q}, \\
				\mu_{{\rm f},n}^{(s)}\leftarrow\tilde{\mu}_{{\rm f}, n}^{(s)}+\frac{\pi^{(s-1)}-1}{\pi^{(s)}}
				\left(\tilde{\mu}_{{\rm f}, n}^{(s)}-\tilde{\mu}_{{\rm f}, n}^{(s-1)}\right), &  \forall n \in \mathcal{N}, \\
				\mu_{{\rm r},n}^{(s)}\leftarrow\tilde{\mu}_{{\rm r}, n}^{(s)}+\frac{\pi^{(s-1)}-1}{\pi^{(s)}}
				\left(\tilde{\mu}_{{\rm r}, n}^{(s)}-\tilde{\mu}_{{\rm r}, n}^{(s-1)}\right), & \forall n \in \mathcal{N}, \\
				\mu_{\rm c}^{(s)}\leftarrow\tilde{\mu}_{\rm c}^{(s)}+\frac{\pi^{(s-1)}-1}{\pi^{(s)}}
				\left(\tilde{\mu}_{\rm c}^{(s)}-\tilde{\mu}_{\rm c}^{(s-1)}\right), & 
			\end{array}
			\right.
		\end{equation}
	\end{subequations}
	where $\tilde{\mu}_{{\rm m}, k}^{(s)}-\tilde{\mu}_{{\rm m}, k}^{(s-1)}$, 
	$\tilde{\mu}_{{\rm b}, k}^{(s)}-\tilde{\mu}_{{\rm b}, k}^{(s-1)}$,
	$\tilde{\mu}_{{\rm e}, q}^{(s)}-\tilde{\mu}_{{\rm e}, q}^{(s-1)}$, 
	$\tilde{\mu}_{{\rm f}, n}^{(s)}-\tilde{\mu}_{{\rm f}, n}^{(s-1)}$,
	$\tilde{\mu}_{{\rm r}, n}^{(s)}-\tilde{\mu}_{{\rm r}, n}^{(s-1)}$ and
	$\tilde{\mu}_{\rm c}^{(s)}-\tilde{\mu}_{\rm c}^{(s-1)}$ are Nesterov momentums.  Also, $\pi^{(s)}$ is a particularly tuned parameter satisfying \cite{beck2009fast}
	\begin{align}\label{La_fac_AA_3}
		\pi^{(0)}=1,\ \pi^{(s)}=\frac{1}{2}\left(1+\sqrt{1+4\left(\pi^{(s-1)}\right)^{2}}\right).
	\end{align}
	
	\begin{figure}[!t]
		\includegraphics[width=3.5in]{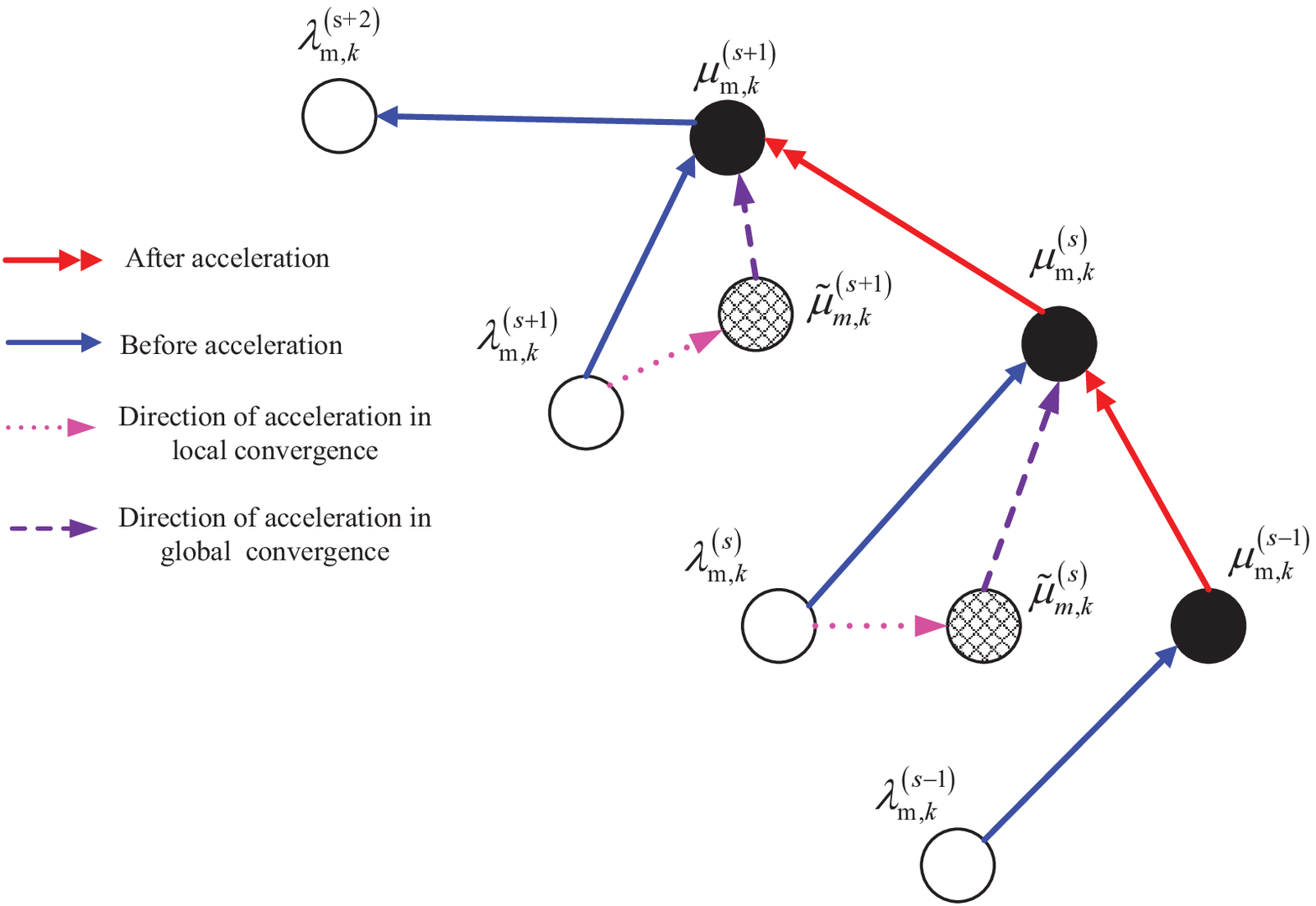}
		\centering{}
		\vspace{-5pt}
		\caption{An illustration of joint Nesterov and heavy-ball acceleration.}
		\label{Fig_2}
	\end{figure}
	
	The great insight of the acceleration lies in the heavy-ball and Nesterov momentums (without these momentums, the accelerated algorithm would reduce to Algorithm~\ref{Algorithm1}). In particular, both momentums use previous updates to generate an overshoot, so that the updates using \eqref{La_fac_AA_1} and \eqref{La_fac_AA_2} is more aggressive than the conventional gradient step \eqref{La_fac} in Algorithm~\ref{Algorithm1}. For illustration purposes,  by using the contour iteration diagram, Fig.~\ref{Fig_2} shows the advantages of the accelerated algorithm, where the update of $\lambda_{{\rm m},k}$ is taken for instance. On the other hand, to ensure the accelerated algorithm not miss the optimal point, it is necessary to control these momentums via a sequence of monotonically increasing parameters $\left\{\pi^{(s)}\right\}$ \cite{10.5555}.  With $\left\{\pi^{(s)}\right\}$ updated according to \eqref{La_fac_AA_3}, the accelerated algorithm is guaranteed to converge to the global optimum of \eqref{P_t1} at a rate of $\mathcal{O}(1/s^{2})$ \cite{9222312}. As the other steps of the accelerated algorithm are identical to Algorithm~\ref{Algorithm1} except for the projected gradient step, the detailed algorithm procedure is omitted for brevity.

	\section{First-Order Algorithm for Finding Feasible Initial Point}\label{Sec_ffp}
	
	In the preceeding section, we have developed Algorithm 1 and its accelerated algorithm for solving \eqref{P0} provided that an initial feasible point is available. In practice, however, it is challenging to find a feasible point for the nonconvex constrained problem \eqref{P0}. To address this issue, the feasibility problem is transformed into an equivalent nonconvex optimization with two simple and independent sets of constraints. Then, by designing another first-order algorithm, we solve the equivalent problem in the SCA framework via stochastic subgradient descent.
	
	\subsection{Problem Transformation}
	By virtue of \eqref{R_ck}-\eqref{R_gk} and \eqref{C_fhh}, the feasibility problem of \eqref{P0} can be written as 
	\begin{equation}\label{fea}
		\text{find}\ \mathcal{V},\ {\rm s.t.} \ \eqref{C2}, \eqref{C3}, \eqref{C4}, \eqref{C_fhh}.
	\end{equation}
	To determine the (in)feasibility of \eqref{fea}, it can be equivalently transformed into the following optimization problem with only two simple sets of constraints \cite{8003316}:
	\begin{subequations} \label{fea_11}
		\begin{align}
			\min_{\mathcal{V}}\ h(\mathcal{V})\triangleq
			&\sum_{n \in \mathcal{N}}\Big(\sum_{j\in \bar{\cal G}}\mathds{I}\{\left\|\bm{A}_{n}\bm{V}_{j}\|_{F}^{2}\right\}{\cal R}_{j}-R_{{\rm F},n}(\mathcal{V})\Big)^{+}\nonumber\\
			&{}+\sum_{q\in{\cal Q}} \left(\mathcal{F}^{-1}\left(e_{q}\right)-P_{{\rm R}, q}(\mathcal{V})\right)^{+}, \label{fea_1} \\
			{\rm s.t.} \ 
			& \sum_{j\in0\cup{\cal G}}\|\bm{A}_{n}\bm{V}_{j}\|_{F}^{2} \leq p_{n}, \ \forall n \in \mathcal{N}, \label{fea_c}\\
			&\sum_{n\in{\cal N}}\|\bm{U}_{n}\|_{F}^{2}\leq p_{c},\label{fea_cc}
		\vspace{-5pt}
		\end{align}
	\end{subequations}
	where the objective function $h(\mathcal{V})$ consists of the sum of $N+Q$ nonconvex functions, with each measuring the degrees of violating its corresponding constraint via a hinge-loss function. If an optimal solution ${\cal V}^{*}$ of \eqref{fea} can be obtained for which $h(\mathcal{V}^{*})=0$, then ${\cal V}^{*}$ is feasible for \eqref{P0}. Otherwise, \eqref{P0} is infeasible and from the values of $Q+N$ loss components, we can determine which constraints cause infeasibility. Nonetheless, as $h(\mathcal{V})$ given by \eqref{fea_11} is nonconvex and NP-hard, solving problem \eqref{fea_11} remains a challenging task. To proceed, in the next subsection the original problem \eqref{fea_11} is approximated as a sequence of convex upper bounds with SCA techniques, then a first-order algorithm is developed to solve the approximated problem via stochastic subgradient descent.
	
	\subsection{First-order Algorithm for Solving \eqref{fea_11}}
	To tackle the nonconvexity of $h({\cal V})$ in \eqref{fea_1}, we construct a set of convex upper bounds by using \eqref{C_ffh222}, \eqref{rate_ff} and \eqref{phi_ee1}:
	\begin{align}
		h(\mathcal{V})
		&\leq \sum_{q\in{\cal Q}} \underbrace{\left(\mathcal{F}^{-1}\left(e_{q}\right)-\bar{P}_{{\rm R}, q}^{(t)}(\mathcal{V})\right)^{+}}_{\hbar_{{\rm e}, q}^{(t)}(\mathcal{V})}\nonumber\\ 
		&\quad{}+ \sum_{n \in \mathcal{N}}\underbrace{\left(\hat{R}_{{\rm A},n}^{(t)}({\cal V})-\bar{R}_{{\rm F},n}^{(t)}(\mathcal{V})\right)^{+}}_{\hbar_{{\rm f}, n}^{(t)}(\mathcal{V})}\triangleq \bar{h}^{(t)}(\mathcal{V}). \label{fea_2}
	\vspace{-5pt}
	\end{align}
	With \eqref{fea_2}, the $t^{\rm th}$ SCA subproblem of \eqref{fea_11} can be rewritten as
	\begin{equation} \label{fea_3}
		\min_{\mathcal{V}} \bar{h}^{(t)}(\mathcal{V}), \text{ s.t. } \eqref{fea_c}, \eqref{fea_cc}.
	\end{equation}
	Since the constraints \eqref{fea_c} and \eqref{fea_cc} are convex and $\bar{h}^{(t)}(\mathcal{V})$ is a smooth and Lipschitz continuous convex function, a straightforward method to solve \eqref{fea_3} is the subgradient descent method, updated by
	\begin{equation}\label{fea_sd}
		{\cal V}^{(s+1)} = \prod_{\cal X}\left({\cal V}^{(s)}-\bar{\nu}_{s}\nabla_{\cal V}\bar{h}^{(t)}(\mathcal{V})\right), 
	\vspace{-5pt}\end{equation}
	where $\prod_{\cal X}$ denotes the Euclidean projection operator onto ${\cal X}$, with ${\cal X}$ being the set of constraints \eqref{fea_c} and \eqref{fea_cc}; $\bar{\nu}_{s}$ is the step size at the $s^{\rm th}$ iteration, and $\nabla_{\cal V}\bar{h}^{(t)}(\mathcal{V})$ is the subgradient with respect to ${\cal  V}$. However, since this subgradient descent requires computing $N+Q$ subgradients at each step, it becomes extremely intensive for large $N$ and $Q$. 
	
	To reduce computation burden, stochastic subgradient descent, where an index $m_{s}$ is randomly drawn from a uniform distribution defined on the index set ${\cal M}=\{1, \cdots, N, \cdots, N+Q\}$ at each iteration $s$ and then perform the update:
	\begin{equation}\label{fea_ssd}
		{\cal V}^{(s+1)} = \prod_{\cal X}\left({\cal V}^{(s)}-\bar{\nu}_{s}\nabla_{\cal V}\hbar_{m_s}^{(t)}(\mathcal{V})\right), 
		\vspace{-5pt}
	\end{equation}                                                                                                            
	where $\nabla_{\cal V}\hbar_{m_s}^{(t)}(\mathcal{V})$ is subgradient drawn from the subgradients set of $\Big\{\{\hbar_{{\rm e}, q}^{(t)}(\mathcal{V})\}_{q\in{\cal Q}},\{\hbar_{{\rm f}, n}^{(t)}(\mathcal{V})\}_{n\in{\cal N}}\Big\}$ with respect to $\mathcal{V}$. The advantage of stochastic gradient descent is that the updates are computationally cheaper than the subgradient descent at each iteration, as it only needs to compute the gradient of a single component function.
	Specifically, $\left\{\left\{ {\bm V}_{j} \right\}_{j \in 0\cup{\cal G}},\left\{ {\bm U}_{n}\right\}_{n \in \mathcal{N}} \right\}$ is first used to compute $\left\{ \left\{ {\bm Z}_{{\rm v},j} \right\}_{j\in 0\cup{\cal G}}, \left\{ {\bm Z}_{{\rm u},n}\right\}_{n \in \mathcal{N}} \right\}$ as per
	\begin{subequations} \label{fea_33}
		\begin{align}
			\bm{Z}_{{\rm v},j} 
			&= {\bm V}_{j}-\bar{\nu}_{s}\nabla_{{\bm V}_{j}}\hbar_{{\rm v},m_{s}}^{(t)}(\mathcal{V}), \ \forall j \in 0\cup{\cal G}, \\
			\bm{Z}_{{\rm u},n} 
			&= {\bm U}_{n}-\bar{\nu}_{s}\nabla_{{\bm U}_{n}}\hbar_{{\rm u},m_{s}}^{(t)}(\mathcal{V}), \ \forall n \in \mathcal{N},
		\end{align}
	\end{subequations}
	where  $\nabla_{{\bm U}_{n}}\hbar_{{\rm u},m_{s}}^{(t)}(\mathcal{V})$ and $\nabla_{{\bm V}_{j}}\hbar_{{\rm v},m_{s}}^{(t)}(\mathcal{V})$ are subgradients drawn from the sets of  $\Big\{\{\nabla_{{\bm U}_{n}}\hbar_{{\rm f}, n}^{(t)}(\mathcal{V})\}_{n\in{\cal N}}\Big\}$ and $\Big\{\{\nabla_{{\bm V}_{j}}\hbar_{{\rm e}, q}^{(t)}(\mathcal{V})\}_{q\in{\cal Q}},\{\nabla_{{\bm V}_{j}}\hbar_{{\rm f}, n}^{(t)}(\mathcal{V})\}_{n\in{\cal N}}\Big\}$ detailed in Appendix~\ref{Appendix-D}, respectively. Then, to satisfy the constraints \eqref{fea_c} and \eqref{fea_cc},  $\left\{ \left\{ {\bm Z}_{{\rm v},j}\right\} _{j \in 0\cup{\cal G}},\left\{ {\bm Z}_{{\rm u},n}\right\}_{n\in {\cal N}}\right\}$ is further projected onto the constraint sets of \eqref{fea_c} and \eqref{fea_cc}, which can be formulated as
	\begin{subequations}\label{fea_44}
		\begin{align}
			\min_{\mathcal{V}}\   
			& \sum_{j \in \bar{\cal G}}\|{\bm{V}}_{j}-\bm{Z}_{{\rm v}, j}\|_{F}^{2}
			+ \sum_{n \in \mathcal{N}}\|{\bm{U}}_{n}-\bm{Z}_{{\rm u}, n}\|_{F}^{2}, \label{fea_4} \\
			{\rm s.t.}\ & \eqref{fea_c}, \eqref{fea_cc}.
		\end{align}
	\end{subequations}
	As \eqref{fea_44} is a quadratic programming with only two simple and independent constraints, by recalling the Karush-Kuhn-Tucker (KKT) conditions, its solution can be readily expressed as  
	\begin{subequations} \label{fea_5}
		\begin{align}
			{\bm V}_{j} &= 
			\left\{
			\begin{array}{rl}
				\bm{Z}_{{\rm v}, j}, & \text{if} \ \Lambda_{n} \leq p_{n} \\
				\sum_{n \in \mathcal{N}}\sqrt{\frac{p_{n}}{\Lambda_{n}}}\bm{A}_{n}\bm{Z}_{{\rm v}, j}, & \text{otherwise }
			\end{array}, \right.  \forall j \in 0\cup{\cal G},  \label{fea_5_v} \\
			{\bm U}_{n} &= 
			\left\{
			\begin{array}{rl}
				\bm{Z}_{{\rm u},n}, & \text{if}\ \Lambda_{c} \leq p_{c} \\
				\sum_{n \in \mathcal{N}}\sqrt{\frac{p_{c}}{\Lambda_{c}}}\bm{Z}_{{\rm u}, n}, & \text{otherwise}
			\end{array}, \right.  \forall n \in \mathcal{N}, \label{fea_5_u} 
		\end{align}
	\end{subequations}
	where $\Lambda_{n} \triangleq \sum_{j \in 0\cup{\cal G}}\|\bm{A}_{n}\bm{Z}_{{\rm v}, j}\|_{F}^{2}$ and $\Lambda_{c} \triangleq \sum_{n \in {\cal N}}\|\bm{Z}_{{\rm u}, n}\|_{F}^{2}$.
	
	In summary, the first-order algorithm developed for finding a feasible point of problem \eqref{P0} is formalized in Algorithm~\ref{Algorithm2}. In particular, the transformed problem \eqref{fea_11} is iteratively solved with iteration over $t$ in the SCA framework while each SCA subproblem \eqref{fea_3} is solved with iterations over $s$. As the constraint set of \eqref{fea_c} and \eqref{fea_cc} is closed and convex, with step size $\bar{\nu}_{s}=\mathcal{O}\left(1/{\sqrt{s}}\right)$, the iteration with respect to $s$ is guaranteed to converge to the global minimum of \eqref{fea_3} \cite{7080879}. Moreover, like Algorithm~\ref{Algorithm1}, the iteration with respect to $t$ is guaranteed to converge to a stationary point of \eqref{fea_11}. It is noteworthy that, as Algorithm~\ref{Algorithm2} is also based on the SCA framework, it needs to be initialized from a feasible point of problem \eqref{fea_11}. For simplicity, as shown in Step~1 of Algorithm~\ref{Algorithm2}, a feasible initial point can be readily obtained by equally allocating the respective total Tx power of each RRH and the computation center to each service. Finally, it remains to mention that Algorithm~\ref{Algorithm2} would at least guarantee the convergence to a KKT point of feasibility problem \eqref{fea_11}. While there is no theoretical guarantee for achieving global optimality of \eqref{fea}, SCA-based algorithms were empirically demonstrated to be highly successful in convergence to the global optimal cost $h(\mathcal{V}^{*})=0$ (i.e., attain feasibility) in a finite number of iterations \cite{8003316, 6954488}. Otherwise, we need to run Algorithm~\ref{Algorithm2} starting with different initial points.
	
	\begin{algorithm}[t]
		\small
		\caption{First-Order Algorithm for Finding an Initial Feasible Point of  \eqref{P0}}
		\label{Algorithm2}
		\begin{algorithmic}[1]
			\STATE \textbf{Initialization:} Generate an initial point denoted
			${\bm V}_{j}^{(0)} =\sqrt{\frac{p_{n}}{Md_{\rm a}(G+1)}}\bm{\varDelta}_{MN\times d_{\rm a}}, \forall j \in 0\cup{\cal G}$, 
			and ${\bm U}_{n}^{(0)} = \sqrt{\frac{p_{c}}{NMd_{\rm f}}}\bm{\varDelta}_{M\times d_{\rm f}}, \forall n \in \mathcal{N}$,  where each element  of  $\bm{\varDelta}_{MN\times d_{\rm a}}$ and  $\bm{\varDelta}_{M\times d_{\rm f}}$ is a normalized complex number with its phase uniformly distributed in $(-\pi,\pi]$;
			\STATE {$t = 0$;}
			\REPEAT
			\STATE {Compute $\bm{\varXi}_{{\rm b}, k}^{(t)}$, 
				$\bm{\varXi}_{{\rm f}, n}^{(t)}$
				$\bm{\varXi}_{{\rm m}, k}^{(t)}$, 	
				$\bm{\varUpsilon}_{{\rm b}, k}^{(t)}$, 	
				$\bm{\varUpsilon}_{{\rm f}, n}^{(t)}$,
				$\bm{\varUpsilon}_{{\rm m}, k}^{(t)}$, 	
				and
				$\varphi_{{\rm e}, q}^{(t)}$, $\forall q \in \mathcal{Q}$,
				$\hat{\rho}_{{\rm v},n,j}^{(t)}$,			
				$\hat{\cal R}_{j}$, $\forall j \in 0\cup{\cal G}$,
				$\forall n \in \mathcal{N}$ according to \eqref{rate_ff}-\eqref{rate_uu} and \eqref{phi_ee1};}
			\STATE {$\mathcal{V} = \mathcal{V}^{(t)}$;}
			\REPEAT
			\STATE {Compute $\Big\{\{\nabla_{{\bm V}_{j}}\hbar_{{\rm e}, q}^{(t)}(\mathcal{V})\}_{q\in{\cal Q}},\{\nabla_{{\bm V}_{j}}\hbar_{{\rm f}, n}^{(t)}(\mathcal{V})\}_{n\in{\cal N}}\Big\}$ and $\Big\{\{\nabla_{{\bm U}_{n}}\hbar_{{\rm f}, n}^{(t)}(\mathcal{V})\}_{n\in{\cal N}}\Big\}$ as per \eqref{E_p_1}-\eqref{E_p_2} and \eqref{E_r};}
			\STATE {Update $\{ \bm{Z}_{{\rm v}, j}\}_{j \in 0\cup{\cal G}}$ and $\{ \bm{Z}_{{\rm u}, n}\}_{n \in \mathcal{N}}$ as per \eqref{fea_33};}
			\STATE {Update $\left\{ {\bm V}_{j}\right\}_{j \in 0\cup{\cal G}}$ and $\{ \bm{U}_{n}\}_{n \in \mathcal{N}}$ according to \eqref{fea_5};}
			\UNTIL {convergence}
			\STATE {Output $\mathcal{V}^{(t+1)} = \mathcal{V}$;}
			\STATE {$t \leftarrow t+1$;}
			\UNTIL {convergence}
		\end{algorithmic}
	\end{algorithm}	\vspace{-10pt}
	
	\section{Numerical Results and Discussions}\label{Sec_sim}
	In this section, Monte-Carlo simulation experiments are carried out to evaluate the proposed first-order algorithms for joint beamforming of wireless fronthaul and access links in ultra-dense C-RANs with SWIPT. All experiments are performed on MATLAB R2019b running on a Windows X64 machine with 3.7 GHz CPU and 32 GB RAM. In the simulation setup, the RRHs and IUs/EUs are randomly and uniformly distributed over a coverage area of $300 \times 300\ \text{m}^2$. The computation center is located at the center of this area with $L=16$ antennas. The channel model of access and fronthaul links are similar to the model in \cite{8283646}. The main system parameters used in our simulation experiments are summarized in Table~\ref{Table-I}. For simplicity, we assume all group services have the same priority (i.e., $\alpha_{j}=1, \forall j \in 0\cup{\cal G}$); the minimal energy harvested requirements for each EU are identical (i.e., $e_{q} = e_{\min}, \forall q \in \mathcal{Q}$); and each RRH are limited by the same allowable maximal Tx power (i.e., $p_{n} = p_{\rm r}, \forall n \in {\cal N}$).
	
	The step size $\nu_{s}$ used in Algorithm~\ref{Algorithm1} is fixed to unity and the step size $\bar{\nu}_s$ in Algorithm~\ref{Algorithm2} is set to $2/(M\sqrt{s})$. The parameters in \eqref{obj_up} are fixed as $\rho_{1}=10^{5}$ and $\rho_{2}=10^{4}$.  The iteration of either algorithm terminates if the relative change of the corresponding objective function between two consecutive iterations is less than $10^{-4}$. All the simulation results below are obtained by making an average over $10^{3}$ simulation trials. 
	
	\begin{table*}[!t]
		\caption{Simulation Parameter Setting}
		\vspace{-5pt}
		\label{Table-I}
		\centering
		\begin{tabular}{l l !{\vrule width1.2pt} l l}
			\Xhline{1.2pt}
			\bf{Parameter} & \bf{Value} &  \bf{Parameter} & \bf{Value}  \\
			\Xhline{1.2pt}
			Noise power for (user, RRH) & (-94, -102) dBm & 	Antenna Gain for (computation center, RRH)	& (9, 0) dBi \\
			Number of antennas at (IU, EU)  & (2, 2) & 			Number of data streams for (fronthaul, access) links & (2, 2)\\      	      	
			Number of (IU, EU) & (9, 4) & 						Number of antennas at each RRH  & 4 \\
			Number of groups  &3 & 								Energy conversion coefficients & (116, 2.3)\\
			The maximum harvested power	& 37.5 mW & 			The sensitivity of energy harvester  & 0.08 mW\\
			\Xhline{1.2pt}
		\end{tabular}
	   \vspace{-10pt}
	\end{table*}

	\vspace{-5pt}\subsection{Complexity Analysis and Comparison}
	As per Algorithm~\ref{Algorithm1}, the complexity of updating ${\bm V}_{j}^{\lozenge}$ and ${\bm U}_{n}^{\lozenge}$ is $\mathcal{O}(M^{3}N^{3}d_{\rm a}^{3}(G+1)+NL^{3}d_{\rm f}^{3})$ at each iteration \cite{8322450}, and the complexity to update the dual variables is $\mathcal{O}({(2N+2K+Q+1)}/\epsilon_{2})$, where $\epsilon_{1}$ is the target convergence accuracy of inner iterations. Therefore, the overall complexity of solving \eqref{P0} via Algorithm~\ref{Algorithm1} is $\mathcal{O}(T_{\max1}(M^{3}N^{3}d_{\rm a}^{3}(G+1)+NL^{3}d_{\rm f}^{3}+(2N+2K+Q+1)/\epsilon_1))$, where $T_{\max1}$ is the number of outer iterations to converge. Likewise, the computational complexity of the accelerated Algorithm 1 can be computed as $\mathcal{O}(T_{\max1}(M^{3}N^{3}d_{\rm a}^{3}(G+1)+NL^{3}d_{\rm f}^{3}+(2N+2K+Q+1)/\sqrt{\epsilon_1}))$.
	
	By using a similar approach as above, the complexity of Algorithm~\ref{Algorithm2} can be computed and given by $\mathcal{O}(T_{\max2}(MNd_{\rm a}(G+1)+NLd_{\rm f}+(G+N+1)/{\epsilon_2}))$, where $\epsilon_{2}$ is the target convergence accuracy of inner iterations and $T_{\max2}$ is the number of outer iterations to converge. 
	
	For comparison purposes, as summarized in Table~\ref{Table-II}, three traditional second-order algorithms are accounted for in the simulation experiments:
	\begin{itemize}
		\item[1)] Interior-point method (IPM), where each SCA subproblem \eqref{P_t1} is solved by an IPM solver, e.g., MOSEK, yielding  complexity $\mathcal{O}(T_{\max1}((2K+N+Q)MNd_{\rm a}(G+1)+NLd_{\rm f})^{3.5})$ \cite{7120191}. 
		\item[2)] Finding a feasible point via IPM (FP-IPM), where each convex subproblem \eqref{fea_3}  is solved via an IPM solver, yielding complexity $\mathcal{O}(T_{\max2}(N+Q)(MNd_{\rm a}(G+1)+NLd_{\rm f})^{3.5})$ \cite{NOUM_1}.
		\item[3)] Optimal branch-and-reduce-and-bound  (BRB) algorithm, which is capable of finding the global optimal solution of \eqref{P0} by using monotonic optimization, yet with extremely high complexity $\mathcal{O}(T_{\max3}(N+Q)(MNd_{\rm a}(G+1)+NLd_{\rm f})^{3.5})$, where $T_{\max3}$ is the number of iterations for the ``BRB'' algorithm to converge and it is very large if a predetermined desired accuracy $\epsilon$ is small \cite[Eq. (30)]{NOUM_1}. 
	\end{itemize}
	
	\begin{table*}[!t]
		\caption{Comparison of Computational Complexity}
		\vspace{-5pt}
		\label{Table-II}
		\centering
		\begin{tabular}{lll}
			\Xhline{1.2pt}
			\textbf{Algorithm} & \textbf{Computational Complexity}  \\
			\Xhline{1.2pt}
			Algorithm~\ref{Algorithm1}  & $\mathcal{O}(T_{\max1}(M^{3}N^{3}d_{\rm a}^{3}(G+1)+NL^{3}d_{\rm f}^{3}+(2N+2K+Q+1)/\epsilon_1))$  \\
			Accelerated Algorithm~\ref{Algorithm1}  & $\mathcal{O}(T_{\max1}(M^{3}N^{3}d_{\rm a}^{3}(G+1)+NL^{3}d_{\rm f}^{3}+(2N+2K+Q+1)/\sqrt{\epsilon_1}))$  \\
			Algorithm~\ref{Algorithm2}  & $\mathcal{O}(T_{\max2}(MNd_{\rm a}(G+1)+NLd_{\rm f}+(G+N+1)/{\epsilon_2}))$  \\
			IPM & $\mathcal{O}(T_{\max1}((2K+N+Q)MNd_{\rm a}(G+1)+NLd_{\rm f})^{3.5})$ \\
			FP-IPM & $\mathcal{O}(T_{\max2}(N+Q)(MNd_{\rm a}(G+1)+NLd_{\rm f})^{3.5})$\\
			BRB & $\mathcal{O}(T_{\max3}(N+Q)(MNd_{\rm a}(G+1)+NLd_{\rm f})^{3.5})$ \\
			\Xhline{1.2pt}
		\end{tabular}
		\vspace{-10pt}	
	\end{table*}

	\vspace{-10pt}
	\subsection{Convergence Behavior of Proposed Algorithms}
	
	\begin{figure}[!t]
		\includegraphics[width=3.0in]{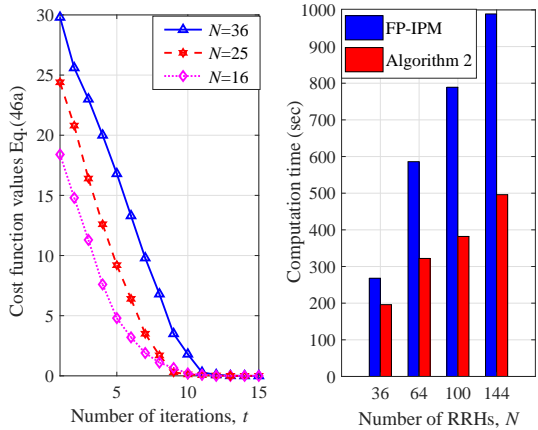}
		\centering{}
		\vspace{-5pt}
		\caption{The convergence of Algorithm~\ref{Algorithm2} for finding the initial feasible point: (left) the learning curves; (right) the computation time versus the number of RRHs  ($p_{\rm r} = 30 \text{ dBm}$, $p_{\rm c} = 40 \text{ dBm}$, and $e_{\min} = 2 \text{ mW}$). }
		\label{Fig_3}
		\vspace{-15pt}
	\end{figure}
	
	Figure~\ref{Fig_3} illustrates the convergence behavior of Algorithm~\ref{Algorithm2} for finding an initial feasible point of problem \eqref{P0}, where the left panel depicts the value of the cost function $h({\cal V})$ given by \eqref{fea_11} versus the number of SCA iterations. It is shown that the values of \eqref{fea_1} decrease to zero after about $14$ SCA iterations under different number of RRHs (i.e., $N = 16, 25, 36$). That is, Algorithm~\ref{Algorithm2} is capable of quickly finding an initial feasible point of \eqref{P0}. On the other hand, the right panel of Fig.~\ref{Fig_3} shows the computation time versus $N$. It is seen that, compared with the ``FP-IPM'' method, Algorithm~\ref{Algorithm2} can greatly reduce the computation time, and the reduciton of computation time becomes more evident as $N$ increases. This demonstrates that the proposed Algorithm~\ref{Algorithm2} is more suitable for ultra-dense C-RANs than the traditional second-order approaches. 

	\begin{figure}[!t]
	\includegraphics[width=3.0in]{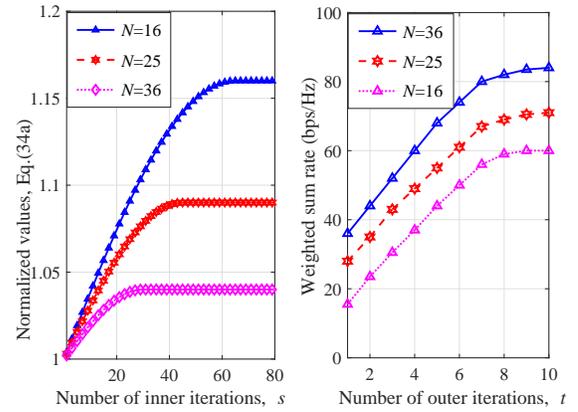}
	\centering{}
	\vspace{-5pt}
	\caption{The convergence of the two-layer iterations inherent in Algorithm~\ref{Algorithm1}: (left) the convergence behavior of the inner iteration; (right) the convergence behavior of the outer iteration  ($p_{\rm r} = 30 \text{ dBm}$, $p_{\rm c} = 40 \text{ dBm}$, and $e_{\min} = 2 \text{ mW}$). }
	\label{Fig_4}
	\vspace{-15pt}
	\end{figure}
	
	Figure~\ref{Fig_4} shows the convergence behavior of the two-layer iterations inherent in Algorithm~\ref{Algorithm1}, where the simulation setting is identical to that in Fig.~\ref{Fig_3}. As observed from the left panel of Fig.~\ref{Fig_4}, the inner iteration converges after about $60$ iterations in the case of $N = 36$, and the convergence speed goes faster as $N$ decreases. On the other hand, the right panel of  Fig. \ref{Fig_4} shows that the outer iteration converges after about $10$ iterations even if $N = 36$. This demonstrates that Algorithm~\ref{Algorithm1} can efficiently find an optimal solution.

	\begin{figure}[!t]
	\includegraphics[width=3.0in]{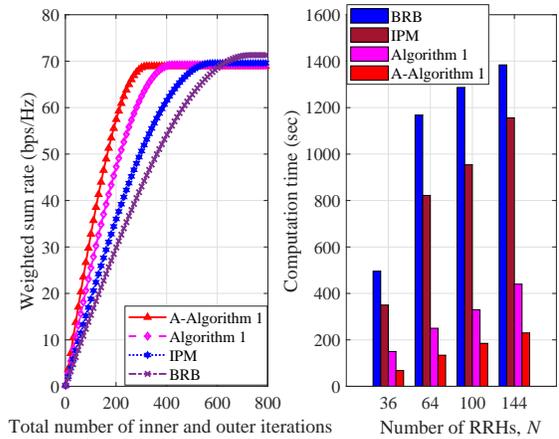}
	\centering{}
	\vspace{-5pt}
	\caption{The performance of Algorithm~\ref{Algorithm1}: (left) the convergence behavior ($N=25$); (right) the computation time versus $N$ ($p_{\rm r} = 30 \text{ dBm}$, $p_{\rm c} = 40 \text{ dBm}$, and $e_{\min} = 2 \text{ mW}$). }
	\label{Fig_5}
	\vspace{-15pt}
	\end{figure}
	
	The left panel of Fig.~\ref{Fig_5} compares the WSR of the proposed first-order algorithms (including Algorithm~\ref{Algorithm1} and its accelerated version (shorted as ``A-Algorithm~1'')), the classic second-order algorithm ``IPM'', and the optimal ``BRB'' algorithm. It is observed that the optimal BRB algorithm obtains the WSR about $71.3$ bps/Hz while the proposed algorithms get about $68.9$ bps/Hz. Precisely speaking, the loss of WSR is only $3.36\%$, which is caused by the smooth approximation \eqref{C_fhh}. By contrast, the left panel of Fig.~\ref{Fig_5} shows that the proposed algorithms converge more than twice as fast as the ``BRB'' algorithm. On the other hand, the right panel of Fig.~\ref{Fig_5} illustrates the computation time of the proposed algorithms, in comparison with that of ``IPM'' and ``BRB''. It is clear that, even though the computation time of each algorithm increases with the number of RRHs (i.e., $N$), the proposed Algorithm~\ref{Algorithm1} has much shorter computation time than the second-order ``IPM'' and ``BRB'' algorithms, and the accelerated algorithm spends the shortest time. Moreover, it is seen from Fig.~\ref{Fig_5} that, as the number of RRHs $N$ increases, the advantages of the proposed Algrithm~1 and its accelerated version become more obvious in terms of computation time. Clearly, this comparison of computation time coincides with the computational complexity analysis in Table~\ref{Table-II}.
	
	\vspace{-10pt}
	\subsection{Performance Comparison}
	In this subsection, the performance of the proposed scheme in this work is compared with two benchmark schemes:
	\begin{itemize}
		\item Benchmark 1: In this scheme \cite{8283646}, the fronthaul and access links are jointly designed but each user is served by only one RRH, unlike our scheme where each user is cooperatively served by multiple RRHs.
		\item Benchmark 2: In this scheme \cite{7858584}, the fronthaul links are firstly designed under the power constraint of computation center and, then, the access links are designed under the constraint of fronthaul capacities, unlike our joint design of fronthaul and access links.
	\end{itemize}

	\begin{figure}[t]
	\includegraphics[width=3.0in]{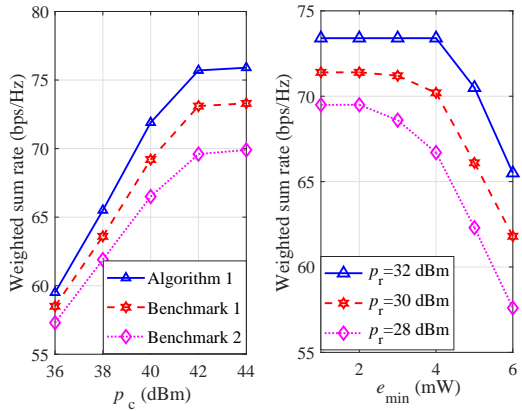}
	\centering{}
	\vspace{-5pt}
	\caption{Comparison of the weighted sum rate: (left) the WSR versus $p_{\rm c}$ ($p_{\rm r}=30 \text{ dBm}$, $e_{\min}=2 \text{ mW}$, and $N=25$); (right) the WSR versus $e_{\min}$ ($p_{\rm c}=40 \text{ dBm}$ and $N=25$).}
	\label{Fig_6}
	\vspace{-15pt}
	\end{figure}
	
	The left panel of Fig.~\ref{Fig_6} depicts the WSR versus the maximal Tx power of computation center (say, $p_{\rm c}$). It is seen that the proposed scheme outperforms the aforementioned benchmark schemes. The reason behind this superiority is that the proposed scheme makes a joint design of the fronthaul and access links by optimizing beamforming and group service association. However, for the ``Benchmark 1'' scheme, although it also jointly optimizes the fronthaul and access links, only unicast transmission is considered and each user is assumed to be associated with only one RRH. As for the ``Benchmark 2'' scheme, the fronthual and access links are designed separately. On the other hand, it is observed that, for all considered schemes, the WSR first increases with $p_{\rm c}$ and then gets saturated. This is because  when $p_{\rm c}$ is large enough, the WSR is dominated by the Tx power of RRHs.
	
	The right panel of Fig.~\ref{Fig_6} illustrates the trade-off between the minimum requirement of harvested energy (say, $e_{\min}$) of each EU and the WSR of information service, obtained by the proposed scheme. It is seen that, for a fixed Tx power of RRH, the WSR keeps constant when $e_{\min}$ is small but it decreases significantly when $e_{\min}$ becomes large. The reason behind this observation is  that if $e_{\min}$ is small, the requirement of harvested energy can be easily satisfied and the WSR is not sensitive to the change of $e_{\min}$. However, when $e_{\min}$ becomes large, more Tx power of RRHs is allocated to fulfill the stringent energy requirement of EU and thus, less power is allocated for data transmission. On the other hand, for a fixed minimum requirement of harvested energy, increasing the Tx power of RRH (say, $p_{\rm r}$) benefits higher WSR, as expected. 
	
	\begin{figure}[t]
		\includegraphics[width=3.0in]{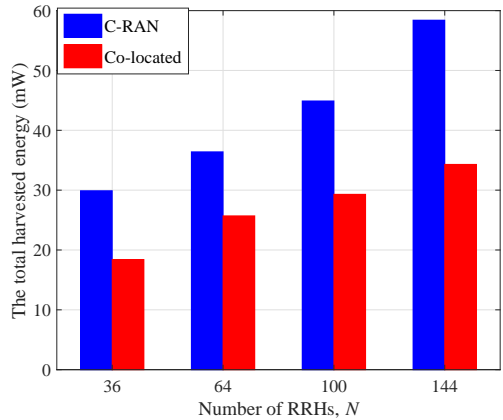}
		\centering{}
		\vspace{-5pt}
		\caption{Comparison of the total harvested energy ($p_{\rm c}=40\ \text{dBm}$, $p_{\rm r}=30 \text{ dBm}$, and $e_{\min}=2 \text{ mW}$).}
		\label{Fig_7}
		\vspace{-15pt}
	\end{figure}
	
	Figure~\ref{Fig_7} illustrates that the total harvest energy versus the numbers of RRHs (i.e., $N$) under different network architectures (``C-RAN'' and ``Co-located''). For the ``Co-located'' architecture, all antennas (the number of antennas is $4N$) are deployed in a central manner. It can be observed that, in two network architectures the total harvested energy increases gradually with the increment of the transmit antennas. Besides, compared to the ``Co-located'' architecture,  the EUs can harvest more energy in the ``C-RAN'' architecture. This is mainly because that, a large portion of radiated power in the ``Co-located'' architecture is used to combat the path loss which emphasizes the benefits of the inherent spatial diversity in distributed antenna systems. These results demonstrate the benefit of network densification, whereby users can be served by RRHs with short distance or potentially with better channel condition.
	
	\section{Concluding Remarks} \label{Sec_con}
	In this paper, we developed a cooperative beamforming for wireless fronthaul and access links in ultra-dense C-RANs with simultaneous wireless information and power transfer, so as to maximize the weighted sum-rate of information services. To efficiently solve this large-scale nonsmooth and nonconvex optimization problem, two low-complexity first-order algorithms were designed for obtaining both the feasible initial point and the final solution. Moreover, to improve the convergence speed, an accelerated algorithm was developed by jointly using the Nesterov and heavy-ball momentums. Simulation results demonstrate that the proposed first-order algorithms achieve almost the same weighted sum-rate as the traditional second-order algorithms. Thanks to the fast convergence speed and low computational complexity, the proposed algorithms are promising for massive access applications, such as massive IoT networks in smart cities. For future work, on account of the limited capacity of fronthaul links in practical C-RANs, it is valuable to design a scalable architecture where multiple computation centers serve disjoint clusters of RRHs. Another critical topic is the collaboration between cloud computing and edge computing, which reduces the transmission delay with mild fronthaul capacity.

	\numberwithin{equation}{section}
	
	\appendices

\section{Proof of Proposition~\ref{Proposition-1}} \label{Appendix-A}
	We start with proving the first property of Proposition~\ref{Proposition-1}. Applying the Woodbury matrix identity to \eqref{Rc_K} gives
	\begin{equation} \label{Rf_2}
		R_{{\rm B},k}\left({\cal V}\right) = \log_2\Big|(\underbrace{\bm{I}_{d_{\rm a}}-\bm{\varTheta}_{{\rm b},k}^{\rm H}\bm{H}_{k}\bm{V}_{0}}_{{\bm Q}_{{\rm b},k}})^{-1}\Big|,
		\vspace{-5pt}
	\end{equation}
	where ${\bm \varTheta}_{{\rm b},k}$ is defined as $\bm{\varTheta}_{{\rm b},k} \triangleq \Big(\sum_{g\in{\cal G}}\bm{H}_{k}\bm{V}_{g}\bm{V}_{g}^{\rm H}\bm{H}_{k}^{\rm H}+\delta_{\rm a}^{2}\bm{I}_{T_{\rm I}}\Big)^{-1}\bm{H}_{k}\bm{V}_{0}$. From \eqref{Rf_2}, it is clear that $R_{{\rm B},k}\left({\cal V}\right)$ is convex over $\bm{Q}_{{\rm b},k}$. Thus, by using the first-order Taylor series expansion at a fixed point $\bm{Q}_{{\rm b},k}^{(t)} = \bm{I}_{d_{\rm a}}-\bm{\varTheta}_{{\rm b},k}^{(t)\rm H}\bm{H}_{k}\bm{V}_{0}^{(t)}$, we have
	\begin{equation} \label{Rf_4}
		R_{{\rm B},k}\left({\cal V}\right)\geq \log_{2}\left|(\bm{Q}_{{\rm b},k}^{(t)})^{-1}\right|-{\rm Tr}\Big((\bm{Q}_{{\rm b},k}^{(t)})^{-1}\bm{Q}_{{\rm b},k}\Big)+d_{\rm a}.
		\vspace{-5pt} 
	\end{equation}
	On the other hand, $\bm{Q}_{{\rm b},k}$ can be majorized by
	\begin{equation} \label{Rf_5}
		\bm{Q}_{{\rm b},k}\preceq\bm{Q}_{{\rm b},k}+\underbrace{(\bm{\varTheta}_{{\rm b},k}-\bm{\varTheta}_{{\rm b},k}^{(t)})^{\rm H}{\bm X}_{{\rm b},k}(\bm{\varTheta}_{{\rm b},k}-\bm{\varTheta}_{{\rm b},k}^{(t)})}_{{\bm E}_{{\rm b},k}}, 
	\end{equation}
	where ${\bm X}_{{\rm b},k} \triangleq \sum_{j\in0\cup{\cal G}}\bm{H}_{k}\bm{V}_{j}\bm{V}_{j}^{\rm H}\bm{H}_{k}^{\rm H}+\delta_{\rm a}^{2}\bm{I}_{T_{\rm I}}$. Substituting \eqref{Rf_5} into \eqref{Rf_4} yields
	\begin{equation} \label{Rf_6}
		R_{{\rm B},k}\left({\cal V}\right)\geq \log_{2}\left|(\bm{Q}_{{\rm b},k}^{(t)})^{-1}\right|-{\rm Tr}\left((\bm{Q}_{{\rm b},k}^{(t)})^{-1}\bm{E}_{{\rm b},k}\right)+d_{\rm a}. 
	\end{equation}
	Then, we show the right-hand-side of \eqref{Rf_6} is equal to $\bar{R}_{{\rm B},k}^{(t)}\left({\cal V}\right)$. By using $\bm{Q}_{{\rm b},k}$ in \eqref{Rf_2} and $\bm{\varTheta}_{{\rm b},k}$ defined after \eqref{Rf_2}, $\bm{E}_{{\rm b},k}$ defined in \eqref{Rf_5} can be rewritten as
	\begin{equation} \label{Rf_7}
		\bm{E}_{{\rm b},k}=\bm{\varTheta}_{{\rm b},k}^{(t){\rm H}}\bm{X}_{{\rm b},k}\bm{\varTheta}_{{\rm b},k}^{(t)}-2{\Re}\{\bm{\varTheta}_{{\rm b},k}^{(t){\rm H}}{\bm H}_{k}\bm{V}_{0}\}+{\bm I}_{d_{\rm a}}.
	\end{equation}
	In view of the expression  $\vartheta_{{\rm b},k}^{(t)}$, $\bm{\varXi}_{{\rm b},k}^{(t)}$, $\bm{\varUpsilon}_{{\rm b},k}^{(t)}$ and $\bm{E}_{{\rm b},k}$, shown in \eqref{rate_f_a}, \eqref{rate_f_D}, \eqref{rate_f_B} and \eqref{Rf_7}, respectively, it is clear that the right-hand-side of \eqref{Rf_6} is equal to $\bar{R}_{{\rm B},k}^{(t)}\left({\cal V}\right)$ and, hence, $R_{{\rm B},k}\left({\cal V}\right)\geq \bar{R}_{{\rm B},k}^{(t)}\left({\cal V}\right)$. Next, we show the equation holds at ${\cal V}={\cal V}^{(t)}$ or equivalently ${\bf V}_j={\bm V}_{j}^{(t)}$. When ${\cal V}={\cal V}^{(t)}$, we have $\bm{\varTheta}_{{\rm b},k}=\bm{\varTheta}_{{\rm b},k}^{(t)}$ and $\bm{Q}_{{\rm b},k}=\bm{Q}_{{\rm b},k}^{(t)}$ and, thus, the equality in \eqref{Rf_4} holds. On the other hand,  \eqref{Rf_5} implies that $\bm{Q}_{{\rm b},k} = \bm{E}_{{\rm b},k}$ if $\bm{\varTheta}_{{\rm b},k} = \bm{\varTheta}_{{\rm b},k}^{(t)}$, and then the equality in \eqref{Rf_6} also holds at ${\cal V}={\cal V}^{(t)}$. Since
	we have shown that the right-hand-side of \eqref{Rf_6} is equal to $\bar{R}_{{\rm B},k}^{(t)}({\cal V})$, it follows that
	$\bar{R}_{{\rm B},k}^{(t)}({\cal V}^{(t)})=R_{{\rm B},k}({\cal V}^{(t)})$.  Similarly, we can prove $\bar{R}_{{\rm M},k}^{(t)}({\cal V})\leq R_{{\rm M},k}({\cal V})$ and $\bar{R}_{{\rm M},k}^{(t)}({\cal V}^{(t)})=R_{{\rm M},k}({\cal V}^{(t)})$.
	
	Next, we prove the second property of Proposition 1. Since the right-hand-side of \eqref{Rf_4} is the first-order expansion of $R_{{\rm B},k}({\cal V})$ at $\bm{Q}_{{\rm b},k}=\bm{Q}_{{\rm b},k}^{(t)}$, we have
	\begin{equation}
		\frac{\partial R_{{\rm B},k}({\cal V}^{(t)})}{\partial \upsilon}=-{\rm Tr}\left((\bm{Q}_{{\rm b},k}^{(t)})^{-1}\frac{\partial \bm{Q}_{{\rm b},k}}{\partial \upsilon}\Big|_{{\cal V}={\cal V}^{(t)}}\right),\label{Rf_8}
	\end{equation}
	where $\upsilon$ represents any element of ${\cal V}$. From \eqref{Rf_5}, we have
	\begin{align}
		\frac{\partial {\bm E}_{{\rm b},k}}{\partial \upsilon}\Big|_{{\cal V}={\cal V}^{(t)}}&=
		\frac{\partial (\bm{\varTheta}_{{\rm b},k}-\bm{\varTheta}_{{\rm b},k}^{(t)})^{\rm H}{\bm X}_{{\rm b},k}}{\partial \upsilon}\Big|_{{\cal V}={\cal V}^{(t)}}(\bm{\varTheta}_{{\rm b},k}-\bm{\varTheta}_{{\rm b},k}^{(t)})\nonumber\\
		&{}+(\bm{\varTheta}_{{\rm b},k}-\bm{\varTheta}_{{\rm b},k}^{(t)})^{\rm H}{\bm X}_{{\rm b},k}\frac{\partial (\bm{\varTheta}_{{\rm b},k}-\bm{\varTheta}_{{\rm b},k}^{(t)})}{\partial \upsilon}\Big|_{{\cal V}={\cal V}^{(t)}}\nonumber\\
		&{}+\frac{\partial {\bm Q}_{{\rm b},k}}{\partial \upsilon}\Big|_{{\cal V}={\cal V}^{(t)}}=\frac{\partial {\bm Q}_{{\rm b},k}}{\partial \upsilon}\Big|_{{\cal V}={\cal V}^{(t)}}.\label{Rf_9}
	\end{align}
	Inserting \eqref{Rf_9} into \eqref{Rf_8} yields $\frac{\partial R_{{\rm B},k}({\cal V}^{(t)})}{\partial \upsilon}=\frac{\partial \bar{R}_{{\rm B},k}^{(t)}({\cal V}^{(t)})}{\partial \upsilon}$, where the equality comes from the fact that the right-hand-side of \eqref{Rf_6} equals $\bar{R}_{{\rm B},k}^{(t)}({\cal V}^{(t)})$. In the same vein, we can derive $\nabla\bar{R}_{{\rm M},k}^{(t)}({\cal V}^{(t)}) = \nabla R_{{\rm M},k}({\cal V}^{(t)})$. This completes the proof.

\section{Proof of Proposition 3} \label{Appendix-B}
	The dual function of \eqref{P_0_up} is defined as \eqref{2.1}
	\begin{align}\label{2.1}
		\mathcal{D}(\mathcal{L}) 
		& \triangleq \min_{\mathcal{V},\mathcal{R}}\ \mathcal{M}\left(\mathcal{V},\mathcal{R},\mathcal{L}\right)\nonumber\\
		& =\min_{\mathcal{V},\mathcal{R}}\ \sum_{j\in0\cup{\cal G}}\rho_{1}\|\bm{V}_{j}-\bm{V}_{j}^{(t)}\|_{F}^{2}+\sum_{q\in{\cal Q}}\lambda_{{\rm e},q}\Big({\cal F}^{-1}\left(e_{q}\right)\nonumber\\
		&{}-\bar{P}_{{\rm R},q}^{(t)}({\cal V})\Big)-\sum_{k\in{\cal K}}\Big[\lambda_{{\rm m},k}\bar{R}_{{\rm M},k}^{(t)}({\cal V})+\lambda_{{\rm b},k}\bar{R}_{{\rm B},k}^{(t)}({\cal V})\Big]\nonumber\\
		&{}+\sum_{n\in{\cal N}}\Big[\lambda_{{\rm f},n}\Big(\hat{R}_{{\rm A},n}({\cal V})-\bar{R}_{{\rm F},n}^{(t)}({\cal V})\Big)+\lambda_{{\rm r},n}\Big(P_{{\rm T},n}\left({\cal V}\right)\nonumber\\
		&{}-p_{n}\Big)+\rho_{2}\|\bm{U}_{n}-\bm{U}_{n}^{(t)}\|_{F}^{2}+\lambda_{\rm c}\big\|\bm{U}_{n}\|_{F}^{2}\Big]+\Big(\sum_{k\in{\cal K}}\lambda_{{\rm b},k}\nonumber\\
		&{}-\alpha_{0}\Big){\cal R}_{0}+\sum_{g\in{\cal G}}\Big(\sum_{k\in{\cal K}_{g}}\lambda_{{\rm m},k}-\alpha_{g}\Big){\cal R}_{g}-\lambda_{\rm c}p_{c},
	\end{align}
	where $\mathcal{M}\left(\mathcal{V},\mathcal{R},\mathcal{L}\right)$ is the Lagrangian function of problem \eqref{P_0_up}. Since $\mathcal{M}\left(\mathcal{V},\mathcal{R},\mathcal{L}\right)$ in \eqref{2.1} is strongly convex over $\mathcal{V}$ and linear over $\mathcal{R}$, the minimum of $\mathcal{M}\left(\mathcal{V},\mathcal{R},\mathcal{L}\right)$ over $\mathcal{V, R}$ is finite if 
	$\sum_{k\in\mathcal{K}_{g}}\lambda_{{\rm m},k}-\alpha_{g}=0, \forall g \in \mathcal{G}$ and $\sum_{k\in\mathcal{K}}\lambda_{{\rm b},k}-\alpha_{0}=0$. 
	Otherwise, we have four other cases, i.e., 
	$\sum_{k \in \mathcal{K}_{g}}\lambda_{{\rm m},k}-\alpha_{g}>0\left(\text{or} <0\right)$ and $\sum_{k \in \mathcal{K}}\lambda_{{\rm b},k}-\alpha_{0}>0\left(\text{or} <0\right)$. 
	Since there are no constraints on $\mathcal{R}_{g}$ and $\mathcal{R}_{0}$ in \eqref{2.1},  $\mathcal{D}(\mathcal{L})$ would be infinite, if $\mathcal{R}_{g}$ (and either or $\mathcal{R}_{0}$) is infinite and $\sum_{k \in \mathcal{K}_{g}}\lambda_{{\rm m},k}-\alpha_{g}\neq0$ (and either or $\sum_{k\in\mathcal{K}}\lambda_{{\rm b},k}-\alpha_{0}=0$). 
	Therefore, $\mathcal{D}(\mathcal{L})$ is finite if and only if
	$\sum_{k \in \mathcal{K}_{g}}\lambda_{{\rm m},k}-\alpha_{g}=0,\forall g \in \mathcal{G}$ and $\sum_{k\in\mathcal{K}}\lambda_{{\rm b},k}-\alpha_{0}=0$. 
	Since the domain of $\mathcal{D}(\mathcal{L})$ is defined as the constraint set over $\mathcal{L}$ so that $\mathcal{D}(\mathcal{L})$ is finite, together with the non-negativity of the dual variables, we get the domain of $\mathcal{D}(\mathcal{L})$ shown in \eqref{dom}.
	
	Then, inserting $\bar{R}^{(t)}_{{\rm B},k}(\mathcal{V})$, $\bar{R}^{(t)}_{{\rm M},k}(\mathcal{V})$, $\hat{R}_{{\rm A},n}^{(t)}({\cal V})$, $\bar{P}^{(t)}_{{\rm R},q}(\mathcal{V})$ and $\bar{R}^{(t)}_{{\rm F},n}(\mathcal{V})$ into \eqref{2.1}, and accounting for that $\sum_{k \in \mathcal{K}_{g}}\lambda_{{\rm m},k}-\alpha_{g}=0$ and $\sum_{k\in\mathcal{K}}\lambda_{{\rm b},k}-\alpha_{0}=0$, as well as dropping the constants independent of $\mathcal{V}$, problem \eqref{2.1} becomes
	\begin{align}\label{2.2}
		\!\min_{\cal V}\ & \rho_{1}\sum_{j\in 0\cup{\cal G}}\|\bm{V}_{j}-\bm{V}_{j}^{(t)}\|_{F}^{2}+\sum_{n\in {\cal N}}\Big(\rho_{2}\|\bm{U}_{n}-\bm{U}_{n}^{(t)}\|_{F}^{2}+\lambda_{\rm c}\!\nonumber\\
		&{}\times\|\bm{U}_{n}\|_{F}^{2}\Big)-2\sum_{q\in{\cal Q}}\lambda_{{\rm e},q}\sum_{j\in0\cup{{\cal G}}}{\Re}\left\{{\rm Tr}\left(\bm{V}_{j}^{(t)\rm H}\bm{F}_{q}^{\rm H}\bm{F}_{q}\bm{V}_{j}\right)\right\}\nonumber\\
		&{}+\sum_{k\in{\cal K}}\Bigg[\lambda_{{\rm m},k}\bigg(\sum_{g\in{\cal G}}{\rm Tr}({\bm V}_{g}^{\rm H}{\bm \varXi}_{{\rm m},k}^{(t)}{\bm V}_{g})+{\Re}\{{\rm Tr}({\bm \varUpsilon}_{{\rm m},k}^{(t)}{\bm V}_{g_{k}})\}\bigg)\nonumber\\
		&{}+\lambda_{{\rm b},k}\bigg({\Re}\{{\rm Tr}({\bm \varUpsilon}_{{\rm b},k}^{(t)}{\bm V}_{0})\}+\sum_{j\in0\cup{\cal G}}{\rm Tr}({\bm V}_{j}^{\rm H}{\bm \varXi}_{{\rm b},k}^{(t)}{\bm V}_{j})\bigg)\Bigg]\nonumber\\
		&{}+\sum_{n\in{\cal N}}\lambda_{{\rm f},n}\Big({\Re}\{{\rm Tr}({\bm \varUpsilon}_{{\rm f},n}^{(t){\rm H}}\bm{U}_{n})\}+\sum_{\ell\in{\cal N}}{\rm Tr}(\bm{U}_{\ell}^{\rm H}{\bm \varXi}_{{\rm f},n}^{(t)}\bm{U}_{\ell})\Big)\nonumber\\
		&{}+\sum_{n\in{\cal N}}\sum_{j\in 0\cup{\cal G}}(\lambda_{{\rm f},n}\hat{\rho}_{{\rm v},n,j}^{(t)}\hat{\cal R}_{j}+\lambda_{{\rm r},n})\|\bm{A}_{n}\bm{V}_{j}\|_{F}^{2}.
	\end{align}
	Since the objective function in \eqref{2.2} is strongly convex over ${\cal V}$, by setting their gradients to zeros over $\left\{\bm{U}_{n}\right\}_{n\in{\cal N}}$ and $\left\{\bm{V}_{j}\right\}_{j\in0\cup{\cal G}}$, the minimizers ${\cal V}^{\lozenge}$ are uniquely given by \eqref{dp_m1}-\eqref{dp_m3}. Finally, substituting the optimal solution $\mathcal{V}^{\lozenge}$ into \eqref{2.1} gives the dual function $\mathcal{D}(\mathcal{L})$ expressed in \eqref{dual}.

\section{Closed-form solution to Problem~\eqref{pro_1}} \label{Appendix-C}
	Since the constraints $\lambda_{\rm c}, \lambda_{{\rm e},q}, \lambda_{{\rm r},n} , \lambda_{{\rm f},n} \geq 0$ in \eqref{dom} are independent of the other constraints, their optimal solutions can be given by $\mu_{\rm c}^{+}$, $\mu_{{\rm e},q}^{+}$, $\mu_{{\rm r},n}^{+}$ and $\mu_{{\rm f},n}^{+}$, respectively, as shown in \eqref{pro_23-a}-\eqref{pro_2}. Thus, problem \eqref{pro_1} reduces to $G$ subproblems over  $\{\lambda_{{\rm m},k}\}_{k\in{\cal K}_{g}}, \forall g \in {\cal G}$, with each expressed as
	\begin{subequations}\label{C2.1}
		\begin{align}
			\min_{\left\{\lambda_{{\rm m},k}\right\} _{k \in \mathcal{K}_{g}}} 
			& \ \sum_{k \in \mathcal{K}_{g}}\left(\lambda_{{\rm m},k}
			-\mu_{{\rm m},k}\right)^{2}, \label{C1.1} \\
			\text{s.t.} 
			& \ \lambda_{{\rm m},k}\geq0,\ \forall k \in \mathcal{K}_{g}, \label{C1.2}\\
			&\ \sum_{k\in\mathcal{K}_{g}}\lambda_{{\rm m},k}-\alpha_{g}=0, \label{C1.33}
		\end{align}
	\end{subequations}
	and one subproblem over $\{\lambda_{{\rm b},k}\}_{k\in{\cal K}}$, expressed as
	\begin{subequations}\label{C2.1c}
		\begin{align}
			\min_{\left\{\lambda_{{\rm b},k}\right\} _{k \in \mathcal{K}}} 
			& \ \sum_{k \in \mathcal{K}}\left(\lambda_{{\rm b},k}
			-\mu_{{\rm b},k}\right)^{2},\label{C1.1c}\\
			\text{s.t.}
			& \ \lambda_{{\rm b},k}\geq0,\ \forall k \in \mathcal{K}, \label{C1.2c}\\
			&\ \sum_{k\in\mathcal{K}}\lambda_{{\rm b},k}-\alpha_{0}=0. \label{C1.33c}
		\end{align}
	\end{subequations} 

	From \eqref{C2.1}, it is not hard to see that  $\lambda_{{\rm m},k}$  can be either zero or positive. If they are positive, they must satisfy the following KKT conditions:
	\begin{align}\label{C1.3}
		\lambda_{{\rm m},k}-\mu_{{\rm m},k}+\frac{\varpi_{g}}{2}=0, \ \forall k \in \mathcal{K}_{g}
	\end{align}
	where $\varpi_{g}$ is the dual variable corresponding to \eqref{C1.33}. Together with the case of $\lambda_{{\rm m},k}=0$, we obtain the optimal solutions of  $\lambda_{{\rm m},k}$, given by \eqref{pro_22}. Finally, substituting \eqref{pro_22} into \eqref{C1.33} yields $\sum_{k \in \mathcal{K}_{g}}\left(\mu_{{\rm m},k}-\frac{\varpi_{g}}{2}\right)^{+}=\alpha_{g}$, from which the value of $\varpi_{g}$ can be determined by using the bisection method. Likewise, we can also attain the optimal solutions \eqref{pro_23} of problem \eqref{C2.1c}, satisfying $\sum_{k \in \mathcal{K}}\left(\mu_{{\rm b},k}-\frac{\varpi_{0}}{2}\right)^{+}=\alpha_{0}$.

\section{Derivations of subgradient} \label{Appendix-D}
	As $\hbar_{{\rm f},\ell}^{(t)}(\mathcal{V})$ is a function of ${\bm V}_{j}$ and ${\bm U}_{n}$ but $\hbar_{{\rm e},q}^{(t)}(\mathcal{V})$ is a function of only ${\bm V}_{j}$, the subgradient of  $\hbar_{{\rm e}, q}^{(t)}({\cal V})$ and $\hbar_{{\rm f},\ell}^{(t)}({\cal V})$ with respect to ${\bm V}_{j}$  can be computed as
		\begin{align}
			\nabla_{{\bm V}_{j}}\hbar_{{\rm e},q}^{(t)}(\mathcal{V}) &=
			\begin{cases}
				\begin{array}{rl}
					-2\bm{F}_{q}^{\rm H}\bm{F}_{q}\bm{V}_{j}^{(t)}, & \text{if } \bar{P}_{{\rm R},q}^{(t)}({\cal V}) \leq {\cal F}^{-1}\left(e_{q}\right); \\
					\bm{0}, & \text{otherwise},
				\end{array}
			\end{cases}\label{E_p_1}\\
			\nabla_{{\bm V}_{j}}\hbar_{{\rm f},\ell}^{(t)}(\mathcal{V}) &=
			\begin{cases}
				\begin{array}{rl}
					2\hat{\rho}_{{\rm v},\ell,j}^{(t)}\hat{\cal R}_{j}\bm{A}_{\ell}\bm{V}_{j}, & \text{if } \hat{R}_{{\rm A}, \ell}^{(t)}({\cal V}) \geq \bar{R}_{{\rm F},\ell}^{(t)}({\cal V}); \\
					\bm{0}, & \text{otherwise}.
				\end{array}
			\end{cases}\label{E_p_2}
		\end{align}
	Then, following similar derivations, we obtain the subgradient of $\hbar_{{\rm f},\ell}^{(t)}(\mathcal{V})$ with respect to  ${\bm U}_{n}$, given by
	\begin{align} \label{E_r}
		\nabla_{{\bm U}_{n}}\hbar_{{\rm f},\ell}^{(t)}(\mathcal{V}) &=
		\begin{cases}
		\begin{array}{rl}
		{\bm \kappa}_{n}, & \text{if}\  \hat{R}_{{\rm A},\ell}^{(t)}({\cal V}) \geq \bar{R}_{{\rm F},\ell}^{(t)}({\cal V}),\ell=n,\\
		{\bm \kappa}_{n,\ell}, & \text{if}\ \hat{R}_{{\rm A},\ell}^{(t)}({\cal V}) \geq \bar{R}_{{\rm F},\ell}^{(t)}({\cal V}),\ell\in{\cal N}\setminus n,\\
		\bm{0}, & \text{otherelse},
		\end{array} 
		\end{cases}
	\end{align}
where ${\bm \kappa}_{n} \triangleq 2\bm{\varXi}_{{\rm f},n}^{(t)}\bm{U}_{n}+\bm{\varUpsilon}_{{\rm f},n}^{(t){\rm H}}$ and ${\bm \kappa}_{n,\ell} \triangleq 2\bm{\varXi}_{{\rm f},\ell}^{(t)}\bm{U}_{n}$.

	\bibliographystyle{IEEEtran}
	\bibliography{References}
	
\begin{IEEEbiography}[{\includegraphics[width=1in,height=1.25in,clip,keepaspectratio]
		{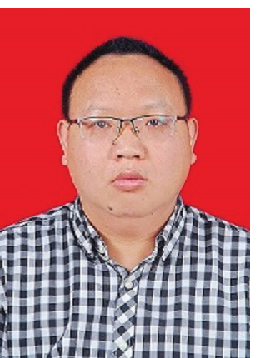}}]{Fangqing Tan}
	received the M.S. degree in communication and information system from Chongqing University of Post and Telecommunications in 2012. He received the Ph.D. degree from Beijing University of Post and Telecommunications in 2017, Beijing, China. From July 2017 to September 2018, he was a Lecturer with Guilin University of Electronic Technology, Guilin, China. He is now a Postdoctoral Fellow at the School of Electronics and Information Technology, Sun Yat-sen University, Guangzhou, China. His research interests  mainly focus on wireless power transfer, multiple antennas communications, and Internet of Things.
	
	He was recognized as an Examplary Reviewer by IEEE WIRELESS COMMUNICATIONS LETTERS in 2020.
\end{IEEEbiography}	

\begin{IEEEbiography}[{\includegraphics[width=1in,height=1.25in,clip,keepaspectratio]
		{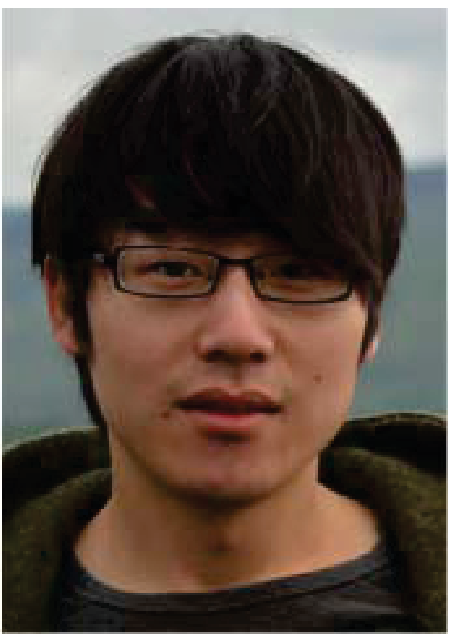}}]{Peiran Wu} (M'16) 
	received the Ph.D. degree in electrical and computer engineering at the University of British Columbia (UBC), Vancouver, Canada, in 2015. From October 2015 to December 2016, he was a Postdoctoral Fellow at the same university. In summer 2014, he was a Visiting Scholar at the Institute for Digital Communications, Friedrich-Alexander-University Erlangen-Nuremberg (FAU), Erlangen, Germany. Since February 2017, he has been with the Sun Yat-sen University, Guangzhou, China, where he is now an Associate Professor. Since 2019, he has been an Adjunct Associate Professor with the Southern Marine Science and Engineering Guangdong Laboratory, Zhuhai, China. His research interests include mobile edge computing, wireless power transfer, and energy-efficient wireless communications. 
	
	He was the recipient of the Fourth-Year Fellowship in 2010, the C. L. Wang Memorial Fellowship in 2011, Graduate Support Initiative (GSI) Award in 2014 from the UBC, German Academic Exchange Service (DAAD) Scholarship in 2014, and the Chinese Government Award for Outstanding Self-Financed Students Abroad in 2014. 
\end{IEEEbiography}

\begin{IEEEbiography}[{\includegraphics[width=1in,height=1.25in,clip,keepaspectratio]
		{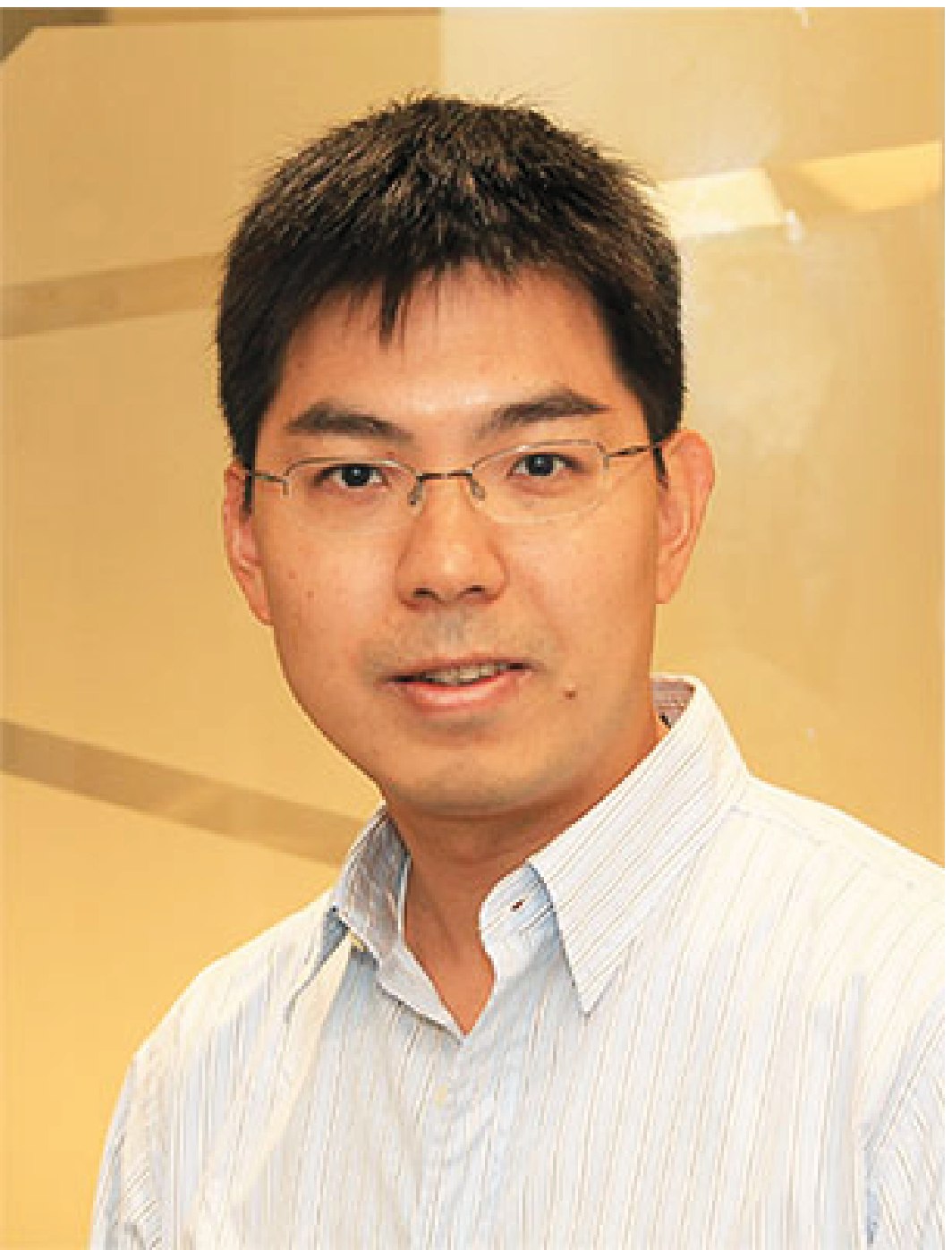}}]{Yik-Chunk Wu} (S'99-M'05-SM'14) 	
	received the B.Eng. (EEE) degree in 1998 and the M.Phil. degree in 2001 from the University of Hong Kong (HKU), and Ph.D. degree from Texas A\&M University in 2005.  From 2005 to 2006, he was with the Thomson Corporate Research, Princeton, NJ, as a Member of Technical Staff.  
	
	Since 2006, he has been with HKU, currently as an Associate Professor.  He was a visiting scholar at Princeton University, in summers of 2015 and 2017.  His research interests are in general areas of signal processing, machine learning and communication systems.  
	
	Dr. Wu served as an editor for {\scshape IEEE Communications Letters} and {\scshape IEEE Transactions on Communications}.  He is currently an associate editor for {\scshape IEEE Transactions on Signal Processing} and an editor for {\scshape Journal of Communications and Networks}. 
\end{IEEEbiography}

\begin{IEEEbiography}[{\includegraphics[width=1in,height=1.25in,clip,keepaspectratio]
		{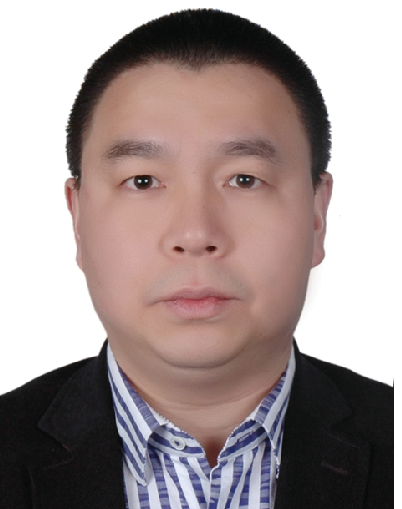}}]{Minghua Xia} (M'12-SM'20) 
	received the Ph.D. degree in Telecommunications and Information Systems from Sun Yat-sen University, Guangzhou, China, in 2007. 
	
	From 2007 to 2009, he was with the Electronics and Telecommunications Research Institute (ETRI) of South Korea, Beijing R\&D Center, Beijing, China, where he worked as a member and then as a senior member of engineering staff. From 2010 to 2014, he was in sequence with The University of Hong Kong, Hong Kong, China; King Abdullah University of Science and Technology, Jeddah, Saudi Arabia; and the Institut National de la Recherche Scientifique (INRS), University of Quebec, Montreal, Canada, as a Postdoctoral Fellow. Since 2015, he has been a Professor with Sun Yat-sen University. Since 2019, he has also been an Adjunct Professor with the Southern Marine Science and Engineering Guangdong Laboratory (Zhuhai). His research interests are in the general areas of wireless communications and signal processing. 
	
\end{IEEEbiography}
\vfill	
	
\end{document}